\documentclass{article}

\usepackage[preprint]{neurips_2025}

\usepackage[utf8]{inputenc} % allow utf-8 input
\usepackage[T1]{fontenc}    % use 8-bit T1 fonts
\usepackage{hyperref}       % hyperlinks
\usepackage{url}            % simple URL typesetting
\usepackage{booktabs}       % professional-quality tables
\usepackage{amsfonts}       % blackboard math symbols
\usepackage{nicefrac}       % compact symbols for 1/2, etc.
\usepackage{microtype}      % microtypography
\usepackage[dvipsnames]{xcolor}        % colors

\newcommand{\systemname}{\emph{HILAD}}

\usepackage{enumitem}
\PassOptionsToPackage{svgnames,table,xcdraw,dvipsnames}{xcolor}
\usepackage[most]{tcolorbox}
\newtcbox{\mybox}[1][gray]{on line, boxsep=2.5pt, boxrule=0pt, left=-1.2pt,right=-1.2pt,top=-1.2pt,bottom=-1.2pt, colback=gray, colframe=gray, arc=0.5mm}
\newtcbox{\myboxx}[1][gray]{on line, boxsep=2.5pt, boxrule=0pt, left=-0.5pt,right=-0.5pt,top=-1pt,bottom=-1pt, colback=gray, colframe=gray,arc=1.4mm}
\newtcbox{\myboxl}[1][Gainsboro]{on line, boxsep=2.5pt, boxrule=0pt, left=-1.2pt,right=-1.2pt,top=-1.2pt,bottom=-1.2pt, colback=Gainsboro, colframe=Gainsboro, arc=0.5mm}

\newtcbox{\myboxxA}[1][gray]{on line, boxsep=2.5pt, boxrule=0pt, left=-0.5pt,right=-0.5pt,top=-0.8pt,bottom=-0.8pt, colback=gray, colframe=gray,arc=1.622mm}  %%This is for text of sec 4

\newtcbox{\mylongbox}[1][gray]{on line, boxsep=2.5pt, boxrule=0pt, left=-1pt,right=-1pt,top=0.8pt,bottom=0.8pt, colback=gray, colframe=gray,arc=2.0mm}  %%This is for small of A1

\newtcbox{\mylongboxT}[1][gray]{on line, boxsep=2.1pt, boxrule=0pt, left=0pt,right=0pt,top=1pt,bottom=1pt, colback=gray, colframe=gray,arc=1.7mm}  %%This is for tniy of A1

\usepackage{amsmath,amsfonts}
\usepackage{algorithm}
\usepackage[noend]{algpseudocode}
\usepackage{array}
\usepackage{stfloats}
\usepackage{url}
\usepackage{graphicx}
\usepackage{multirow}
\usepackage{booktabs} 
\usepackage{orcidlink}
\definecolor{customehighlight}{HTML}{ffffff}
\definecolor{RoyalBlue}{HTML}{4169E1}
\definecolor{Green}{HTML}{008000}

\usepackage{soul}
\sethlcolor{yellow}

\title{A Reliable Framework for Human-in-the-Loop Anomaly
Detection in Time Series}

\author{Ziquan Deng\thanks{Both authors contributed equally to this research.}\\
  University of California, Davis\\
  Davis, USA\\
  \texttt{ziqdeng@ucdavis.edu} \\
  \And
  Xiwei Xuan$^*$\\
  University of California, Davis\\
  Davis, USA\\
  \texttt{xwxuan@ucdavis.edu} \\
  \And
  Kwan-Liu Ma \\
  University of California, Davis\\
  Davis, USA\\
  \texttt{klma@ucdavis.edu} \\
  \And
  Zhaodan Kong\thanks{Corresponding Author.} \\
  University of California, Davis\\
  Davis, USA\\
  \texttt{zdkong@ucdavis.edu} \\
}

\begin{document}

\maketitle

% %%%%%%%%%%%%%%%%%%%%%%%%%%%%% main paper start 1 %%%%%%%%%%%%%%%%%%%%%%%%%%%%%%%

\begin{abstract}

%%%%%%%%%%%%%%%%%%% V1
Time series anomaly detection is a critical machine learning task for numerous applications, such as finance, healthcare, and industrial systems. However, even high-performing models may exhibit potential issues such as biases, leading to unreliable outcomes and misplaced confidence. While model explanation techniques, particularly visual explanations, offer valuable insights by elucidating model attributions of their decision, many limitations still exist---They are primarily instance-based and not scalable across the dataset, and they provide one-directional information from the model to the human side, lacking a mechanism for users to address detected issues. To fulfill these gaps, we introduce \systemname{}, a novel framework designed to foster a dynamic and bidirectional collaboration between humans and AI for enhancing anomaly detection models in time series. Through our visual interface, \systemname{} empowers domain experts to detect, interpret, and correct unexpected model behaviors at scale. Our evaluation through user studies with two models and three time series datasets demonstrates the effectiveness of \systemname{}, which fosters a deeper model understanding, immediate corrective actions, and model reliability enhancement.

% Time series anomaly detection is a critical machine learning task for numerous applications, such as finance, healthcare, and industrial systems. However, even high-performing models may exhibit potential issues such as biases, leading to unreliable outcomes and misplaced confidence. While model explanation techniques, particularly visual explanations, offer valuable insights by elucidating model attributions of their decision, many limitations still exist---They are primarily instance-based and not scalable across the dataset, and they provide one-directional information from the model to the human side, lacking a mechanism for users to address detected issues. To fulfill these gaps, we introduce HILAD, a novel framework designed to foster a dynamic and bidirectional collaboration between humans and AI for enhancing anomaly detection models in time series. Through our visual interface, HILAD empowers domain experts to detect, interpret, and correct unexpected model behaviors at scale. Our evaluation through user studies with two models and three time series datasets demonstrates the effectiveness of HILAD, which fosters a deeper model understanding, immediate corrective actions, and model reliability enhancement.
 
\end{abstract}

\section{Introduction}\label{sec:intro}

Time series anomaly detection is a crucial technique in machine learning (ML), with broad applications such as identifying irregular financial transactions~\cite{ahmed2016survey}, monitoring abnormal health indicators in patient records~\cite{vsabic2021healthcare}, and preempting equipment failures in manufacturing~\cite{li2019mad}. 
Despite its importance, the effectiveness of these models can be compromised by biases---where models make decisions based on incorrect input features, resulting in problematic conclusions. 

\begin{figure}[t!]
\centerline{\includegraphics[width=0.7\columnwidth]{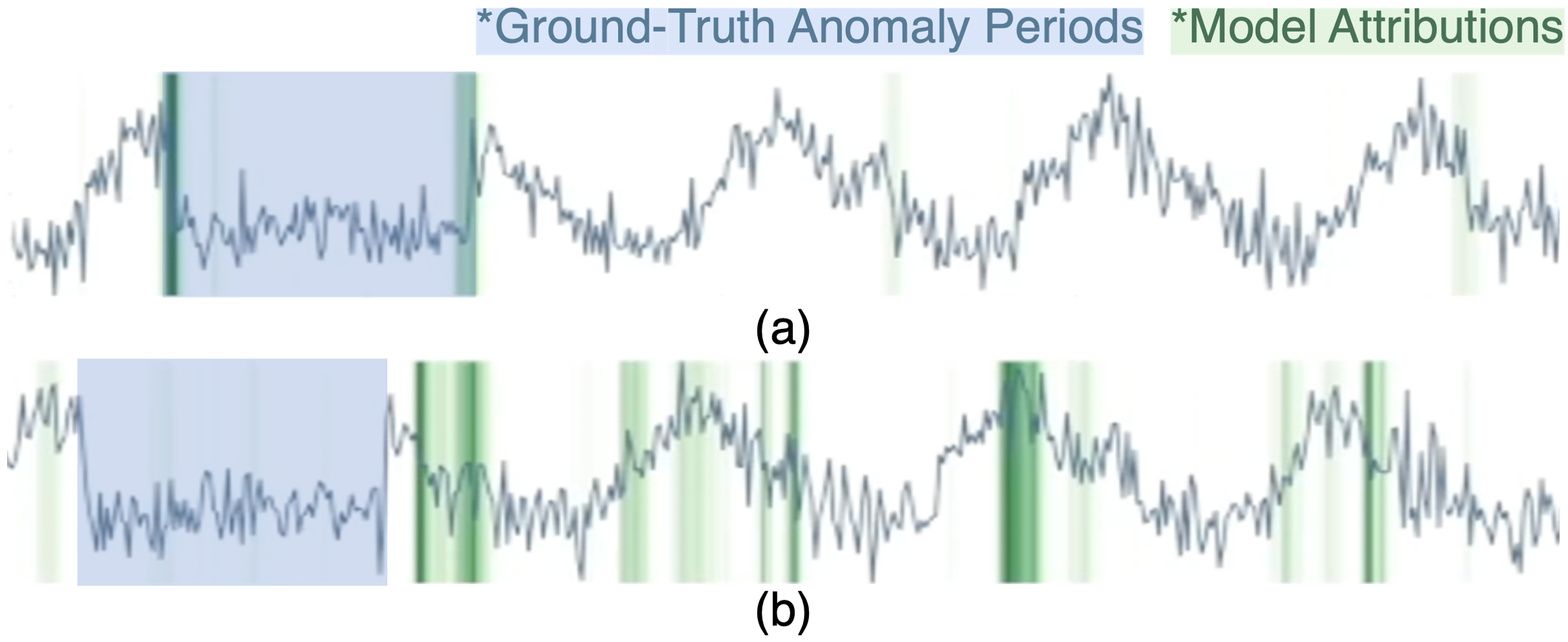}}
\caption{
Two sequences from the NASA MSL dataset with the ground truth label as ``anomaly''. The true time frames corresponding to anomalies are highlighted in {\color{RoyalBlue} blue}, and model's CAM attributes are highlighted in {\color{Green} green}. While the model makes ``anomaly'' prediction for both, regions highlighted by two colors overlap in (a) but not in (b), showing the model leverages wrong features as the anomaly indicator in (b).
}
\label{fig:bias_examples}
\end{figure}

To tackle this issue, explainable AI (XAI) techniques~\cite{dwivedi2023explainable,xuan2024suny} have been introduced to expose the decision logic of anomaly detectors. 
Among them, visual explanation methods like Class Activation Mapping (CAM)~\cite{zhou2016learning,boniol2023dcnn} are particularly effective in attributing importance scores to each time step in a sequence. 
CAM can assign a significance score to each input time step, indicating its contribution to a model prediction. 
% This illustrates the need to move beyond instance-level correctness and address underlying model biases.
For example, Fig.~\ref{fig:bias_examples} illustrates two time sequences with both labeled as ``anomaly'', and the model makes correct predictions. However, CAM reveals a critical distinction: in case (a), the model’s attention (green) aligns well with the actual anomalous segment (blue), while for (b), the attribution of the prediction (green) does not overlap with the ground-truth (blue). 
Although both predicted labels seem correct, only the first reflects proper reasoning. 
This discrepancy demonstrates a hidden model bias, where decisions appear valid but are based on faulty logic, and highlights the need for an explainable system that not only exposes the bias but also helps mitigate such failures.

However, most existing XAI methods are limited in two key aspects: they analyze individual instances without generalizing to systematic model behaviors, and they enable only one-way interpretation—from model to user—without allowing users to intervene or improve the model~\cite{barkouki2023xai}. 
In complex real-world environments, such limitations can lead to brittle deployments, where spurious model logic remains undetected and uncorrected at scale.

To ensure robust and trustworthy anomaly detection, we argue that explanation alone is insufficient. Instead, effective systems must support \emph{human-in-the-loop} workflows where users can not only validate model decisions but also provide corrective feedback that influences model behavior. 
This perspective is supported by recent progress in human-AI collaboration, which shows how visual analytics interfaces can empower users to audit, refine, and debug AI models~\cite{amershi2019guidelines, xuan2025vista, xuan2024attributionscanner}.

In this paper, we introduce \systemname{},  a human-in-the-loop framework with an interactive interface, designed to enhance the reliability of ML models in the context of time series anomaly classification.
Our framework leverages visual explanation techniques to detect model biases, designing attribution-aware clustering and summarization to foster model issue interpretation and validation.
The interface of \systemname{} navigates domain experts through the model validation workflow, facilitating the identification and mitigation of model biases. 
Our spuriousness propagation mechanism allows humans to contribute domain knowledge at scale, where they only need to manually annotate a few clusters, and the estimated spuriousness scores of others are provided automatically for their verification.  
Lastly, we present a method for enhancing the model by rectifying identified errors based on human feedback.
Our bidirectional interaction utilizes human expertise effectively with a feedback loop, allowing humans to enhance AI systems with minimal effort. 

To demonstrate the effectiveness of \systemname{}, we invite domain experts to conduct model enhancement studies with benchmark datasets, and design comparative analyses to quantitatively compare our approach with representative baselines. Our evaluation encompasses both objective metrics and subjective assessments and aims to investigate the following hypotheses:

\begin{enumerate}[start=1,label={[\bfseries H\arabic*]}]
  \item \label{Hypothesis:enhancement} \emph{Human-AI Collaborative Model Enhancement}. 
  The performance of the \systemname{} framework, when aligned with human's domain expertise at the decision level, is significantly superior to conventional automated algorithms that operate on a black-box machine learning model.
  % Human intervention through the \systemname{} visual interface is expected to foster critical insights, thus identifying and rectifying hidden model errors like spurious correlations.
  \item \label{Hypothesis:interpretability} \emph{Human-Centric Model Interpretation}. Participants' subjective insights, in terms of intuitiveness, transparency, and reliability, will be significantly enhanced by direct interaction with models through \systemname{} framework. In contrast, approaches that lack this synergy may face challenges in offering such perceptions. 
\end{enumerate}

\noindent Our contributions are summarized as follows:
\begin{enumerate}[label={\textbullet}]
    \item \systemname{}, a framework that integrates human expertise with ML techniques to enhance the reliability of time series anomaly detection models, allowing for systematically detecting, interpreting, and mitigating model biases.
    
    \item An interactive VA interface allowing for model validation at different scales, supporting effective issue annotation and propagation with minimal but essential human efforts.

    \item A human-AI-collaborative solution for generating actionable insights for model validation and enhancement, which can inspire future research on human-assisted trustworthy AI.
\end{enumerate}

\section{Related Works}
\subsection{Time Series Anomaly Detection Models}\label{subsec:xai}
ML techniques for time series anomaly detection can be broadly categorized into unsupervised and supervised approaches.
When ground-truth anomaly labels are unknown, unsupervised learning techniques are introduced to approximate future data points~\cite{kieu2019outlier,malhotra2015long,taylor2018forecasting,tuli2022tranad}, where the deviations between the predicted and true values serve as anomaly indicators.
On the other hand, if labeled data are available, supervised models can be trained as classifiers to directly predict anomaly labels~\cite{del2021auto,gallicchio2017local,zhao2017convolutional}. Such models can achieve higher accuracy and be utilized in critical areas like fraud detection or medical diagnostics~\cite{nabrawi2023fraud,meena2022bone}. However, supervised models are generally more susceptible to bias issues, causing significant risks in practice~\cite{uddin2019comparing}. In our work, we focus on enhancing supervised models that directly predict anomaly labels, where various architectures such as transformers and CNN~\cite{dosovitskiy2020image,liu2021swin,boniol2023dcnn,liu2021gated,yang2021voice2series} are fully supported, and we aim to detect, understand, and mitigate hidden biases in such models with a human-in-loop approach.

\subsection{Reliability Considerations of Models}

Biased models can cause significant risks in real-world applications~\cite{jung2021time,wang2024use}. 
For instance, if a model identifies normal time series as anomaly indicators, it underscores a fundamental misunderstanding of the underlying anomalous features, leading to potential overlooking of real anomalies~\cite{goswami2022unsupervised, deng2021interpretable}.
While visual explanations such as CAM can reveal such discrepancies, their occurrence raises critical questions~\cite{xuan2024slim} --- Are these discrepancies a common feature across the entire data corpus, or are they isolated incidents? Moreover, once such biases are identified, what are the most effective methods for addressing and mitigating them?
The discussion in Sec.~\ref{sec:intro} points out the limitations of current XAI techniques in fully tackling these challenges. 
This scenario necessitates further research and development of methods that not only highlight but also rectify these biases. 
Enhancing the reliability of anomaly detection models through improved understanding and correction of biases is essential to foster trust and ensure the efficacy of these systems in real-world applications.

\subsection{Human-AI Collaboration}

The growing field of human-AI collaboration has demonstrated the importance of integrating human expertise into machine learning workflows, particularly in high-stakes applications such as medical imaging, finance, and industrial monitoring~\cite{amershi2019guidelines, schaekermann2020expert, xuan2022vac,zhang2023labelvizier,yildirim2024multimodal,xuan2024slim,deng2025anytime}. 
Prior research in interactive AI systems has explored how visual analytics (VA) platforms can support experts in understanding, validating, and improving models, with much of the focus on vision and natural language processing tasks~\cite{yang2022diagnosing,yang2023interactive,xuan2025vista,xuan2024attributionscanner,li2022unified,he2021can,gou2020vatld,liu2012tiara}, as these tasks are generally more interpretable to humans compared to time series data.
For example, VA systems designed in~\cite{xuan2025vista,li2022unified} allow users to iteratively refine and probe deep learning models. 
Similar approaches in object detection and segmentation~\cite{gou2020vatld,he2021can,yan2025vislix} and medical AI~\cite{xuan2024slim,malik2025towards} have enabled human-guided model adjustments. 
These works underscore the potential of bidirectional human-AI collaboration, where users not only interpret model outputs but actively refine them through feedback loops.
In time series anomaly detection, existing approaches~\cite{geiger2020tadgan,alnegheimish2022sintel,liu2022mtv,feng2023timepool} facilitate interactive anomaly investigation by displaying feature attributions and model decisions.
While these methods improve interpretability, they often rely on instance-level explanations, limiting their scalability to larger datasets~\cite{ruta2019sax}.
As a result, users must perform labor-intensive manual inspections, which are impractical in complex industrial or financial environments where anomalies are highly dynamic and demand rapid adaptation.

To address these challenges, we introduce \systemname{} to bridge the gap between visual interpretation and direct model refinement.
\systemname{} supports a bidirectional interaction paradigm, where users can detect, interpret, and correct model misbehaviors at scale. 
Inspired by prior works that leveraged 2D projections for visualizing time series data~\cite{guo2021interpretable,fujiwara2020visual}, we incorporate a projection step to facilitate human interaction. However, unlike directly projecting raw features, we construct an attribution-aware latent space using an aggregated distance matrix over data and model attributions.~\systemname{}~leverages this space as an intermediate structure, supporting scalable annotation propagation and downstream model enhancement, which advances human-AI collaboration beyond prior visual-inspection-based approaches.
By summarizing dominant failure patterns and allowing human adjustments, our system advances the integration of human expertise into AI-driven anomaly detection, enhancing model reliability and interpretability in time series applications.

\section{\systemname{}~Framework}

\begin{figure*}[t!]
\centerline{\includegraphics[width=1\textwidth]{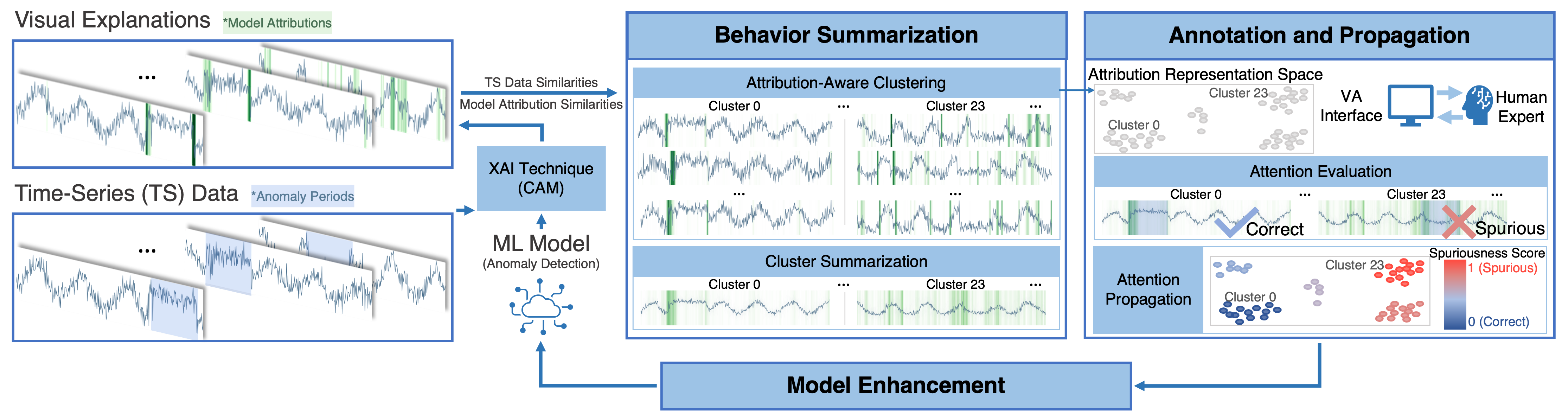}}
\caption{The \systemname{} framework involves three phases: ``Behavior Summarization'', ``Annotation and Propagation'', and``Model Enhancement''. At the ``Behavior Summarization'' phase, we conduct attribution-aware clustering, where each cluster includes TS instances exhibiting similar data features and model attributions. We also visualize the overall pattern of each cluster to streamline user interpretation.
% we conduct attribution-aware clustering to obtain clusters with both similar TS data features and consistent model attributions, and we compute cluster summarization to aggregate each cluster's pattern with a single visualization. 
At the ``Annotation and Propagation'' phase, users interact with our designed VA interface to identify and annotate model bias issues.
These annotations are propagated to unannotated clusters to generate ``Spuriousness scores'' for user verification.
% with the help of Spuriousness propagation.
Lastly, we mitigate the detected issues at the ``Model Enhancement'' phase.}
\label{fig: framework}
\end{figure*}

% \subsection{\systemname{} Framework}

\systemname{}~is a human-in-the-loop framework designed to improve the reliability of anomaly detection models on time series (TS) data. It takes as input a trained anomaly detection model along with TS instances with ground-truth anomaly labels. The system then enables domain experts to interactively inspect, validate, and correct model behaviors through three coordinated phases: ``Behavior Summarization'', ``Annotation and Propagation'', and ``Model Enhancement'', as shown in Fig.~\ref{fig: framework}.

The core idea of \systemname{}~is to assist users in efficiently detecting and addressing model misbehavior caused by spurious attributions, where the model relies on misleading or irrelevant signals to make decisions. To support this, in the first ``Behavior Summarization'' phase, \systemname{}~automatically generates and visualizes clusters with similar data features and model attributions. As illustrated in Fig.~\ref{fig: framework}, each cluster represents a group of TS sequences that share both contextual similarity in terms of data signals and model attribution similarity, and such patterns are visualized in ``cluster summarization'', enabling experts to verify the attribution correctness without inspecting individual sequences. Then, in the following ``Annotation and Propagation'' phase, we enable users to overview clusters' summarized model attributions, validate the exhibition of model issues, and annotate verified issues with a few clicks, where we provide label propagation to facilitate the validation process. Finally, the ``Model Enhancement'' stage systematically addresses these inaccuracies, refining the model's focus to improve its overall performance and reliability.

\subsection{System Interface}
\begin{figure*}[ht!]
\centerline{\includegraphics[width=0.99\textwidth]{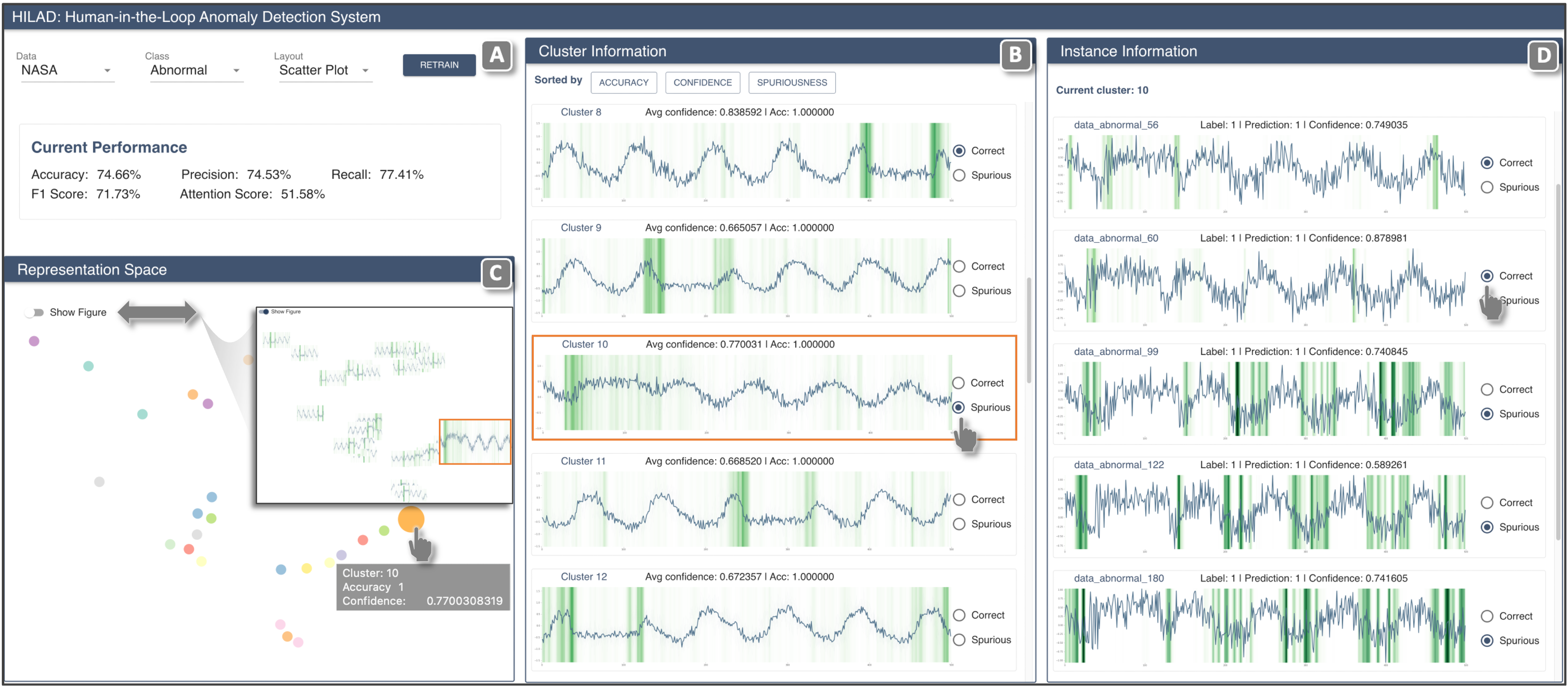}}
\caption{\systemname{} applied to the anomaly detection of a univariate time series classifier trained on the MSL dataset~\cite{hundman2018detecting}. \mybox{\textcolor{white}{A}} System Menu, enabling the selection of dataset, class, and visualization options (Scatter plot
view or Detailed plot view). \mybox{\textcolor{white}{B}} Cluster Information window, showing the data clusters and cluster metrics. \mybox{\textcolor{white}{C}} Representation Space window, showing a visual overview and the relative positions of all data clusters. \mybox{\textcolor{white}{D}} Instance Information window, showing individual time series sequences with model attributions belonging to a selected data cluster.}
\label{fig:Interface}
\end{figure*}

The visual interface of \systemname{} comprises four main components. 
Fig.~\ref{fig:Interface} depicts our interface when validating an anomaly classifier on the Mars Science Laboratory (MSL) dataset~\cite{hundman2018detecting}.

The System Menu (Fig.~\ref{fig:Interface}~\mybox{\textcolor{white}{A}}) provides options for configuration selections, including dataset, class, and visualization layout. 
Our primary objective is to analyze the model's comprehension of anomalous patterns; therefore, we have configured the default class setting to represent anomaly-related instances, with ground-truth or prediction class as an anomaly.
The Cluster Information (Fig.~\ref{fig:Interface}~\mybox{\textcolor{white}{B}}) presents an explainable behavior summarization for the current model in each cluster. 
Each cluster aggregates similar instances that garner analogous attention from the model.
On the top of Cluster Information window, we introduce cluster metrics to facilitate navigation through the dataset. 
The available metrics, including Accuracy, Confidence, and Spuriousness — the latter derived via the Spuriousness Propagation method (Sec.~\ref{sec:spu mitigation}) contingent upon user annotation — are presented to enhance the rapidity of analysis.
Once click/annotate one cluster in Cluster Information window, its corresponding position in 2D space will be highlighted in the Representation Space window (Fig.~\ref{fig:Interface}~\mybox{\textcolor{white}{C}}) and instances in this cluster will be detailed in the Instance Information window (Fig.~\ref{fig:Interface}~\mybox{\textcolor{white}{D}}).
Two visualization layouts are available for Representation Space (Fig.~\ref{fig:Interface}~\mybox{\textcolor{white}{C}}): Scatter plot view or Detailed plot view, allowing for an overarching view of the similarity of clusters in 2D space or a more detailed inspection of each cluster. 
This function is also embedded in Representation Space to facilitate users to switch more flexibly.
We also provide information on the correctness of the current model in terms of classification and attention.

Note that when interacting with the Representation Space window, the other two windows will be updated with coordinate information simultaneously to support user interpretation and issue verification.
Due to the high similarity of adjacent clusters, when users annotate a cluster, they can quickly retrieve their neighbors in this window to speed up the annotation process.
After a cluster is annotated in Cluster Information window, all instances of the cluster in Instance Information window have the same annotation by default. 
To further reduce the error introduced by the clustering algorithm, we allow users to adjust the annotations of individual data in the Instance Information window.
The user must annotate data as correct or spurious at least once.
Upon completing these annotations, the user can initiate model refinement by clicking the Retrain button.
Subsequently, our backend algorithm will fine-tune the model in accordance with the user's annotation, followed by an update to the model's performance metrics shown in System Menu.

\subsection{Behavior Summarization}\label{sec:behavior sum}

At the ``Behavior Summarization'' phase, for each individual instance, we first apply CAM to obtain model attributions correlated to its decision, providing a one-dimensional vector (or mask) with the same length as the time sequence. Each value in this vector falls within the range (0, 1) and reflects the importance of the corresponding timestamp to the model’s prediction. Then, we measure and aggregate similarities between: (1) TS data, according to the Dynamic Time Warping (DTW) Metric, and (2) model attribution masks, according to the cosine similarity, respectively. With this information, we perform attribution-aware clustering to obtain clusters with similar data features and model attributions. To allow for effective user investigation, we visually summarize each cluster to help users identify common patterns of each cluster at a simple glance.

Note that CAM is fully modular and can be replaced with a variety of attribution methods, including Grad-CAM~\cite{selvaraju2020grad}, ScoreCAM~\cite{wang2020score}, and model-agnostic techniques like RISE~\cite{petsiuk2018rise}. Our design does not rely on CAM-specific internals, allowing users to plug in alternative attribution mechanisms depending on the accessibility or specific requirements. We use CAM as our default XAI method, as CAM and CAM-derived techniques remain among the most widely used in time series and other domains due to their simplicity, interpretability, and computational efficiency. And they have been extensively validated across a broad range of models and datasets.

\subsubsection{Time series Context Similarity Calculation}

To measure the similarity between time sequences, we use the prevalent technique, the DTW algorithm,
% In the realm of time series analysis, assessing the similarity between sequences is a pivotal task.
% % , especially when dealing with multi-dimensional and one-dimensional datasets.
% A prevalent technique is the DTW algorithm, 
which strives to minimize the distance between two time-sequenced datasets by dynamically aligning and ``warping" them in the time dimension. A smaller DTW distance indicates a higher similarity.
Formally, for two multivariate time series $T_Q = \{ {T_q}_1, {T_q}_2, \dots , {T_q}_n \}$ and $T_P = \{ {T_p}_1, {T_p}_2, \dots , {T_p}_m \}$, ${T_q}_i \in \mathbb{R}^d$  and ${T_p}_j\in \mathbb{R}^d$ , the DTW distance $d_{DTW}(T_Q, T_P)$ can be computed using the following recursive formula:
\begin{equation}
\begin{split}
   d_{DTW}(T_Q, T_P) = d({T_q}_n, {T_p}_m) \\
   + \min \left\{ 
\begin{array}{l}
DTW({T_q}_{n-1}, {T_p}_m) \\
DTW({T_q}_n, {T_p}_{m-1}) \\
DTW({T_q}_{n-1}, {T_p}_{m-1}),
\end{array}
\right.
\end{split}
\end{equation}
where $d({T_q}_n, {T_p}_m)$ is the Euclidean distance between vectors \({T_q}_n\) and \({T_p}_m\).
The initialization conditions are \(D({T_q}_n, {T_p}_0) = D({T_q}_0, {T_p}_m) = \infty\) and \(D({T_q}_0, {T_p}_0) = 0\).

\subsubsection{Attribution Similarity Calculation}
\label{subsec:atttribution_sim}
Given a model $\mathcal{M}$, trained for TS binary classification task (i.e., normal or abnormal), an input time sequence, $T_Q$, and the predicted class of the $\mathcal{M}$ for input $T_Q$, $c(T_Q)$, the attention score of $T_Q$ is calculated as 
\begin{equation}
    a(T_Q) = CAM(\mathcal{M},T_Q,c(T_Q)),
\end{equation}
where $a(T_Q)$ is the model attribution mask, which is a one-dimensional vector with the same length as $T_Q$.
For input sequences $T_Q$ and $T_P$, the attribution similarity $d_{cos}(a(T_Q), a(T_P))$ is computed by:
\begin{equation}
    d_{cos}(a(T_Q), a(T_P))= 1- \frac{a(T_Q) \cdot a(T_P)}{\|a(T_Q)\|_2 \times \|a(T_P)\|_2},
\end{equation}
where $\|a(T_Q)\|_2$ and $\|a(T_P)\|_2$ denote the L2 norm of vectors $a(T_Q)$ and $T_a(P)$, respectively. For this cosine distance, similar to $d_{DTW}$, a smaller value indicates a higher similarity.

\subsubsection{Attribution-Aware Clustering and Summarization}
To aggregate data with both similar context and similar model attribution, we construct the following aggregated distance $m_{QP}$ to measure the similarity between two instances $T_Q$ and $T_P$:
\begin{equation}
    m_{QP}= \alpha d_{DTW}(T_Q, T_P) + (1-\alpha) d_{cos}(a(T_Q), a(T_P)),
\label{eqn:dist_alpha}
\end{equation}
where $\alpha$ is the weight to balance the data context similarity $d_{DTW}(T_Q, T_P)$ and the model attribution similarity $d_{cos}(a(T_Q), a(T_P))$. Both distances are normalized to the range $[0, 1]$ to ensure scale compatibility and prevent either term from dominating the aggregated distance. 
In this work, we set $\alpha=0.5$.
Thus, for all of the $N$ TS instances, we can get a distance matrix $M_{dist}\in\mathbb{R}^{N\times N}$.
% \begin{equation}
%     M_{dist} = 
%     \begin{bmatrix}
%         m_{1,1} & m_{1,2} &\cdots &m_{1,N}\\
%         m_{2,1} & m_{2,2} &\cdots &m_{2,N}\\
%         \vdots &\vdots &\ddots &\vdots\\
%         m_{N,1} & m_{N,2} &\cdots &m_{N,N}
%     \end{bmatrix}.
% \end{equation}

Finally, we use UMAP~\cite{mcinnes2018umap} to project $M_{dist}$ and apply K-means~\cite{krishna1999genetic} to generate $K$ clusters: 
\begin{equation}
    C = \text{K-means}(UMAP(M_{dist}),K),
\end{equation}
where $C=\{c_1,c_2,\dots,c_K\}$ is the set of clusters, UMAP$(M_{dist})$ is the 2D representation of the data, and $K$ is the desired number of clusters determined by the elbow method. As a follow-up step, we visually summarize the resulting clusters as an integrated visualization for each cluster, which presents the common patterns of time sequences and model attributions in the current cluster, empowering users to grasp model behavior patterns more efficiently.

\subsection{Annotation and Propagation}\label{sec:att eval}

Building upon the aforementioned method, at the ``Annotation and Propagation'' phase, we define a metric termed ``Spuriousness score", $S[c_i]$, to evaluate the presence of spuriousness.
This score ranges from 0 to 1, representing the spurious probability from the least to most likely.
Furthermore, the Label Propagation algorithm~\cite{zhou2003learning} is deployed to automatically estimate probabilities of the unannotated clusters' spuriousness based on users' annotations and neighboring clusters $N(c_i)$.

The representation space is generated by UMAP as described in Sec.~\ref{sec:behavior sum}, where we use the adjacency matrix, $A(i,j)$, to denote the distance between clusters' feature representations of $c_i$ and $c_j$. 
Notably, users only need to provide binary annotation for a cluster, designating it as either ``correct attention" ($S[c_i]=0$) or ``spurious attention" ($S[c_i]=1$). 
Our algorithm (Algo.~\ref{alg:labelprop}, Line 6-14) then propagates this information to compute Spuriousness scores for other clusters. 
Subsequently, these scores are showcased in the Cluster Information window (Fig.~\ref{fig:Interface}~\mybox{\textcolor{white}{B}}).

\begin{algorithm}[t!]
\caption{Label Propagation for Spuriousness Score}
\label{alg:labelprop}
\begin{algorithmic}[1]
\Require
    \State \( C \): Set of clusters.
    \State \( A \): Adjacency matrix of clusters indicating similarity.
    \State \( \mathbf{Y} \): Initial labels, where \( Y[c_i] \in \{0, 1\} \) for labeled clusters and \(-1\) otherwise.
\Ensure
    \State \( \mathbf{S} \): Vector of propagated spuriousness scores for each cluster.
\State Initialize \( \mathbf{S} \gets \mathbf{Y} \)
\For{iteration = 1 to max\_iter}
    \State \( \mathbf{S}^{\text{new}} \gets \mathbf{S} \)
    \For{each cluster \( c_i \in C \)}
        \If{ \( Y[c_i] \neq -1 \) }  %%\Comment{Skip labeled nodes}
            \State \textbf{continue}
        \EndIf
        \State \( \text{WeightedSum} \gets \sum_{c_j \in N(c_i)} A(c_i, c_j) \times S[c_j] \)
        \State \( \text{TotalWeight} \gets \sum_{c_j \in N(c_i)} A(c_i, c_j) \)
        \State \( S^{\text{new}}[c_i] \gets \frac{\text{WeightedSum}}{\text{TotalWeight}} \)
    \EndFor
    \State \( \mathbf{S} \gets \mathbf{S}^{\text{new}} \)
\EndFor
\State \Return \( \mathbf{S} \)
\end{algorithmic}
\end{algorithm}

The ``Spuriousness score" yields two noteworthy advantages. 
Primarily, they facilitate a more streamlined exploration process of clusters, with potentially spurious correlations being conspicuously highlighted. 
This is a pivotal step in aiding users to identify and assess problematic attention. 
Secondly, following users' validation, these scores are leveraged to determine the problematic clusters crucial for issue mitigation, as we elucidate in the following section.

\subsection{Model Enhancement}\label{sec:spu mitigation}

An effective strategy for addressing detected spurious correlations involves model retraining or finetuning, which reduces the model's resilience over potential bias while keeping the architecture intact. 
Specifically, the Core Risk Minimization (CoRM) methodology~\cite{singla2022core} that retrains the model on a noise-corrupted dataset has demonstrated its efficacy in spuriousness mitigation for vision models.
In our work, we adapt CoRM to TS data, which harnesses the masking strategy by injecting random Gaussian noise into the spurious time points, aiming to shift and correct the model's wrong attention by encouraging it to infer decisions with the rest of the data.
% The procedure involves simple steps: first, mask the parts of the data that receive spurious attention, then encourage the model to infer decisions with the rest of the data.
% This idea has been used as a self-supervised pre-training method and spuriousness feature elimination method in Computer Vision (CV) and Natural Language Processing (NLP) fields.
The idea aligns with a data augmentation strategy called jittering for time series data, which can help the model become more robust to variations and be used in masked autoencoders for time series forecasting~\cite{tang2022mtsmae}.

The integration of CoRM into \systemname{}~entails several steps. 
We first extract a set of instances with high spuriousness scores, $T_S$, indicating undesirable attention. 
We also extract a set of instances with the lowest spuriousness, $T_C$, indicating correct attention. 
The attribution masks of $T_S$ highlight spurious regions, which are used to occupy such regions with random Gaussian noise accordingly. 
For an individual time sequence $T_i \in T_S$, this operation can be mathematically represented as  
\begin{equation}
    T^\prime_i = T_i + \mathbf{m} \odot z,
\end{equation}
where $\mathbf{m}$ denotes the CAM mask, $z$ represents the generated Gaussian noise matrix, and $\odot$ denotes the Hadamard product.
% Each variable is of the same dimensions as the input time series data. 
Similarly, for $T_j \in T_C$, we add random Gaussian noise to the non-highlighted areas to enhance the model's attention to abnormal patterns:
\begin{equation}
    T^\prime_j = T_j + (1-\mathbf{m}) \odot z.
\end{equation}

After substituting the original data with these spuriousness-masked and correctness-enhanced data, we proceed to retrain the model. 
Subsequently, an evaluation is conducted to measure the mitigation of spurious correlations. 
In Sec.~\ref{sec:metrics}, we elaborate on the evaluation metrics utilized to quantify the effectiveness of our method in mitigating spurious correlations.

\section{Experimental Design}
In this section, we provide an experimental design to benchmark and evaluate the capabilities of \systemname{}. 
The primary objective of this experiment is to investigate how \systemname{} empowers humans to detect, understand, and correct potential issues in an anomaly classifier for time series data. 

\subsection{Datasets}

\noindent\textbf{Mars Science Laboratory (MSL) Dataset}~\cite{hundman2018detecting} corresponds to the sensor and actuator data for the Mars rover. 
The dataset provides the ground truth anomaly masks that marks the time point when the abnormal behavior occurs.
Since this dataset is known to have many trivial sequences~\cite{wu2021current}, to make a best-faith effort to eliminate mislabeled time points, we build a new dataset based on the two nontrivial ones (\textit{A2} and \textit{A4}). Specifically, we randomly divide the original time series into 1000 equal-length sequences, all of 800 lengths, where the ratio of normal to abnormal data is 3:2.

\noindent\textbf{Simulation Testbed for Exploration Vehicle ECLSS
(STEVE)}~\cite{deng2023causal} is a simplified single-bed $CO_2$ removal system of the Carbon Dioxide Removal Assembly (CDRA) onboard the International Space Station (ISS), at the University of Colorado Boulder. 
A corresponding Simulink model is provided to simulate multiple failure modes similar to the STEVE testbed. 
% For example, a leak at the outlet of the sorbent bed, which may be caused by wear and tear of connectors or human error during maintenance, can be simulated. 
% Air will leak out of the system during adsorption and into the system during desorption due to the pressure difference between the bed and the lab (habitat) environment.
In our study, we conducted a simulation consisting of four cycles, each comprising 80 minutes of $CO_2$ adsorption followed by 80 minutes of desorption. During the third cycle, a simulated leak failure is introduced at a random time point in \textit{Valve 1}, resulting in observable anomalies in three sensors: \textit{Bed Temperature}, \textit{$CO_2$ Concentration}, and \textit{Flow Rate}. These sensor sequences are treated collectively. We generated 1000 sets of sequences with a ratio of 4:1 between normal and abnormal data. Additionally, we marked time points at which abnormal behaviors occurred in the three sensors as ground truth masks.

\noindent\textbf{KPI anomaly detection for AIOps (KPI) dataset}~\cite{li2022constructing} is a real-world, large-scale dataset for evaluating anomaly detection systems in industrial AIOps settings. It contains time series with noisy, high-variance patterns and diverse anomaly types, making it suitable for benchmarking real-world deployment performance. For our experiments, we construct equal-length sequences of 800 time points from four KPI categories with the highest anomaly ratios, resulting in 1,200 time series instances. Additional dataset details are provided in the Appendix.

\subsection{Baselines}

\textbf{Baseline models.} We involve two baseline models in our experiments. One is a fully convolutional network (FCN)~\cite{wang2017time}, which has demonstrated effectiveness on time series classification. The other is TST~\cite{zerveas2021transformer}, a transformer-based model for multivariate time series anomaly detection. These two models have distinct architectures for evaluating our method with various baselines. Note that \systemname{}~is fully applicable to other time series anomaly classification models with different architectures~\cite{boniol2023dcnn,liu2021gated,yang2021voice2series}.
In addition to baseline models, we compare \systemname{}~with the following spuriousness mitigation strategies using data augmentation:

\noindent\textbullet \hspace{5pt}\textit{Random aug.}\hspace{5pt} Randomly select $\Gamma$ instances from the training set for model fine-tuning.

\noindent\textbullet \hspace{5pt}\textit{Random mask.}\hspace{5pt} Select $\Gamma$ instances proposed by our system, but mask random time points.

\noindent\textbullet \hspace{5pt}\textit{Cluster mask.}\hspace{5pt} The $\Gamma$ instances proposed by our system only concede the participant's operation in the Cluster Information window, which means the instance-level operations are discarded. This method uses the same masking strategy as ours.

\subsection{Participants}

% This experiment was approved by the Institutional Review Board at the University of California, Davis.  
This experiment was approved by the Institutional Review Board.
We involve 15 participants in total, of which 13 were male and 2 were female. All participants are graduate students between the ages of 23 to 30 years old. 
All participants passed the test to ensure they could accurately identify and annotate the abnormal pattern.
% All completed the entire experiment on the three datasets. 

\subsection{User Tasks}
In the experiment, participants were asked to interact with the interface of \systemname{} to evaluate a classification model trained on a specific dataset, containing both normal and abnormal time sequences. Specifically, we encouraged them to use as many visual components as possible, follow the system guidance to verify the highlighted issues, interpret them with the provided information, and annotate the confirmed errors within a restricted time. In each experiment, participants were only allowed to perform model retraining once to better quantify the results.

With the awareness of the \systemname{} workflow, participants explored the general model behavior concerning different clusters of data through the Cluster Information window, where they also investigated further details from the Instance Information window that updated simultaneously with their interactions. 
They were asked to mark and annotate detection issues, such as a model using normal time points to determine an abnormal time series, which could be achieved by simply clicking the ``Spurious" button in the Cluster Information window. At this point, all instances corresponding to this cluster were marked as spurious.
Participants could further adjust the spuriousness labels of single instances in the Instance Information window.
Similarly procedure could be conducted to label and adjust a cluster related to correct attention.
As long as one cluster is annotated as spurious or correct, our algorithm will approximate the attention correctness of other clusters according to their similarities in the representation space, resulting in a spuriousness approximation score for every cluster. Participants can click the ``spuriousness" button provided in our interface to rank clusters according to this score, allowing for adjustments and confirmation of the approximated model issues.
When the participants confirmed the first round of issue inspection could be concluded, they would click the retrain button, and our system performed the model retraining based on the participants' feedback.

\subsection{Procedure}
In the following, we introduce the procedure of our user study. 
% The overall practical accomplishment of the experiment took between 30 and 45 minutes.

\subsubsection{Introduction and Preparations}

Before the experiment begins, participants receive detailed instructions explaining the experiment's purpose and overall structure before it begins. Next, they are introduced to the key concepts involved in time series classification tasks, including metrics such as confidence and accuracy, possible abnormal patterns, and spurious model attention. Subsequently, an in-depth orientation to \systemname{} is provided, highlighting its various components and functionalities. This orientation covers the System Menu, Representation Space, Cluster Information, and Instance Information, elucidating their respective roles in detecting anomalies. The primary goal communicated to the participants is to employ \systemname{} for monitoring effectively and analyzing the model's behavior, focusing on its attention mechanisms.

\subsubsection{Familiarization Procedure}

A simplified dataset was introduced to acquaint participants with the functions of the \systemname{} interface. 
This served as a practical tutorial on the tools available in each window. 
During this phase, participants gained insights into the relationships between different clusters and developed skills in identifying and annotating abnormal patterns.
It is important to note that this segment was designed solely for hands-on learning, and its results were not included in the primary evaluation. Once participants demonstrated proficiency in accurately identifying and annotating abnormal patterns, they conducted the experimental run.

\subsubsection{Experimental Run}

In the experimental run, 
% participants were introduced to the \systemname{} with two space domain-related datasets that mimic its real-world use for anomaly detection.
participants' tasks were to critically analyze the behavior of each classification model across three datasets.  
% The main variable in this phase was the dataset itself, and the sequence of dataset presentation varied systematically among participants.     
% This approach aimed to mitigate any potential bias from learning curve effects.
We executed three trials for each dataset to further effectively compare the distinctions between the automated algorithm and the human-AI collaboration method. 
These trials varied in the scope of information provided to the participants: the first trial provided access to only the System Menu window, the second trial included System Menu, Cluster Information, and Representation Space windows, and the third trial offered a comprehensive view with the complete all four windows.
The first trial in our study exemplifies the application of the automated algorithm, while the latter two trials represent the human-AI collaboration methods.
To ensure the integrity of our results and eliminate any potential biases associated with the sequence of exposure, the order of these trials was randomized for each participant. This approach was essential to assess the impact of varying levels of information availability on the performance and decision-making processes in the context of human-AI interaction.

\subsubsection{Postprocessing} 

Once all experimental runs are concluded, participants can reflect on the experiment and share their insights and feedback regarding \systemname{}. 
This qualitative data is gathered through questionnaires designed to assess participants' subjective perceptions of human-AI cooperation, specifically focusing on intuition, transparency, and reliability. 
This data provides valuable insights into the user experience and the system's effectiveness in facilitating anomaly detection, evaluation, and correction.

\subsection{Performance Evaluation Metrics}\label{sec:metrics}
We measure the effectiveness of our proposed framework in terms of classification and attention performance.
From the perspective of classification performance, we use the standard Accuracy, Precision, Recall, and $F_1$ Score to evaluate the performance of our method comprehensively.
% The details of the metrics are listed as follows:
% \begin{gather}\label{eqn:PRF}
%    Acc=\frac{TP+TN}{TP+TN+FP+FN}, P=\frac{TP}{TP+FP}, \notag\\R=\frac{TP}{TP+FN}, F_1=\frac{2\cdot P\cdot R}{P+R},\notag 
% \end{gather}
% where TP, TN, FP, and FN are four classification results, denoting true positive, true negative, false positive, and false negative, respectively.
% $Acc$ is a measure of the overall correctness of a classification model. 
% $P$ represents the ability of the method to distinguish anomalies, and a higher $P$ accurately indicates fewer false anomalies. 
% $R$ is to determine the ability of the method to detect all the anomalies. 
% Since The $P$ and $R$ metrics often appear contradictory, the  $F_1$ score is a metric for comprehensively considering these two indicators and providing a balanced overall performance of the method.
% When the data set has imbalanced classes, F1 Score might be a better metric than accuracy.

In the perspective of attention accuracy performance, we adjust the relevance accuracy ($RA$) proposed in \cite{wickstrom2020uncertainty} to quantify how accurate a model is at locating ground-truth time points at which the abnormal behavior occurs.
The core idea of relevance accuracy is to compare the $k$ most attention time points for a prediction with the ground-truth abnormal time points.
For a given time sequence $T_i$, let $G(T_i)$ denote the set of continuous or discontinuous ground-truth abnormal time points, where $|G(T_i)|$ is the number of ground-truth time points, and $R_i(k)$ denote the $k$ time steps that get the most attention for the prediction of the model.
To apply this metric to data with varying numbers of ground-truth time points, we set $k=(1+M\%)\times |G(T_i)|$ and denote $R_i(k)$ as $R_i(M\%)$ in our work.
In this analysis, the sequence of the ground-truth time points is disregarded, given that each ground-truth time point is treated with equal significance.
Optimal relevance accuracy is attained when all elements of $G(T_i)$ are encompassed within $R_i(M\%)$, ideally with the minimal possible value of $M$.
For a time sequence $T_i$, $RA(T_i)$ can be calculated as:
\begin{equation}
    RA(T_i)=\frac{|R_i(M\%)|\cap G(T_i)}{|G(T_i)|},
\end{equation}
where $\cap$ is the intersection of two sets and $|\cdot|$ is the cardinality of a set. 
The time points the model is attention to are ranked based on CAM value before they are compared to $G(T_i)$.
Therefore, simply highlighting all time points will not result in a high relevance accuracy score. The model needs to highlight some time points as being more important than others.

To assess the participants’ subjective perception of human–machine cooperation, a questionnaire with a five-point Likert Scale~\cite{de2010five} with the following items was used.

% \begin{figure*}[ht!]
% \centerline{\includegraphics[width=1\textwidth]{Figures/Obj1.png}}
% \caption{Tukey HSD post-hoc test results for MSL dataset. The left plot illustrates classification performance, while the right one depicts attention accuracy. Methods not sharing any letter are significantly different by the Tukey-test at the 5\% level of significance.  }
% \label{fig:Obj1}
% \end{figure*}

\subsection{Statistical Analysis}\label{sec:Statistical Ana}

For Hypothesis 1, since all five methods involved in the comparison have been tested an equal number of times, we directly assess their performance using the metrics introduced in Section \ref{sec:metrics}. 
Therefore, our hypothesis tests reduces to:
\begin{gather}\label{eqn:hyp1}
   H_0: {\mu_{Acc}}_1=\dots= {\mu_{Acc}}_5, {\mu_{P}}_1=\dots= {\mu_{P}}_5, \notag
   \\{\mu_{R}}_1=\dots= {\mu_{R}}_5, {\mu_{F_1}}_1=\dots= {\mu_{F_1}}_5, \notag
   \\{\mu_{RA}}_1=\dots= {\mu_{RA}}_5; \notag
      \\H_1: {\mu_{Acc}}_1\neq\dots\neq {\mu_{Acc}}_5, {\mu_{P}}_1\neq\dots\neq {\mu_{P}}_5, \notag
   \\{\mu_{R}}_1\neq\dots\neq {\mu_{R}}_5, {\mu_{F_1}}_1\neq\dots \neq {\mu_{F_1}}_5, \notag
   \\{\mu_{RA}}_1\neq\dots\neq {\mu_{RA}}_5. \notag
\end{gather}

To evaluate participants' subjective perceptions of human-AI cooperation, we administered a questionnaire featuring a five-point Likert scale designed to assess key aspects such as intuition, transparency, and reliability. 
In this subjective evaluation, we compare three methods: (1) Random Augmentation, driven by a purely automatic algorithm. When employing this method, only the System Menu is accessible within our interactive interface, and the user interaction is limited to operating the retrain button; (2) Human-AI Cluster Mask, which, upon execution, presents windows A, B, and C on our interface. Users can assess the correctness of attention by clusters but lack the capability to inspect and adjust individual data points within each cluster; (3) Human-AI \systemname{}, the comprehensive version, where users are allowed to refine their assessments in window D. 
By contrasting (1) with (2) and (3), we aim to elucidate the impact of increasing user engagement on subjective evaluation. Conversely, by contrasting (2) with (3), we seek to discern the added benefits by enabling detailed individual data adjustments within the clusters.

\subsection{Annotation Efficiency Evaluation Settings}
We quantify and compare the human effort between \systemname{}~and multiple representative baselines that also utilize human annotations. Specifically, our comparative study involves three active learning baselines: random sampling, uncertainty-based sampling, and diversity-based sampling. To align with prior practices, we view our basic model as the warm-up model that is trained with labeled data and might have spurious issues. Then, during the human interaction phase, we present the user with individual time sequences with CAM highlighting, and ask them to select from binary options, ``correct’’ or ``spurious’’, indicating whether spurious correlation exists. View D in the \systemname{}~interface is used as the annotation interface. We fix the annotation budget as 160, which is 20\% of the total instances. Additional settings are as follows.
\begin{itemize}[leftmargin=*]
    \item Random-based sampling. Instances are randomly sampled from the dataset and listed in the interface for user annotation. This reflects real-world, unstructured manual debugging workflows.
    \item Uncertainty-based sampling. Individual instances are ranked in descendingly order by model uncertainty (e.g., based on softmax entropy). Users are given instances in order and assign spuriousness labels to each of them.
    \item Diversity-based sampling. We use the basic setting to cluster data features and sample instances nearest to each cluster centroid. Then present the instances to the user in a random order.
\end{itemize}
Considering the practical challenges, we re-invite five of our participants in the original study. We asked them to perform annotations following four settings: ours, random, uncertainty, and diversity. Their orders are random to reduce familiarity-caused noise, and we have a 10-minute break in between. Note that for \systemname{}, the users are asked to determine how many instances to annotate, according to their interactions and observations through our interface.

\section{Results}
\begin{figure*}[t!]
\centerline{\includegraphics[width=1\textwidth]{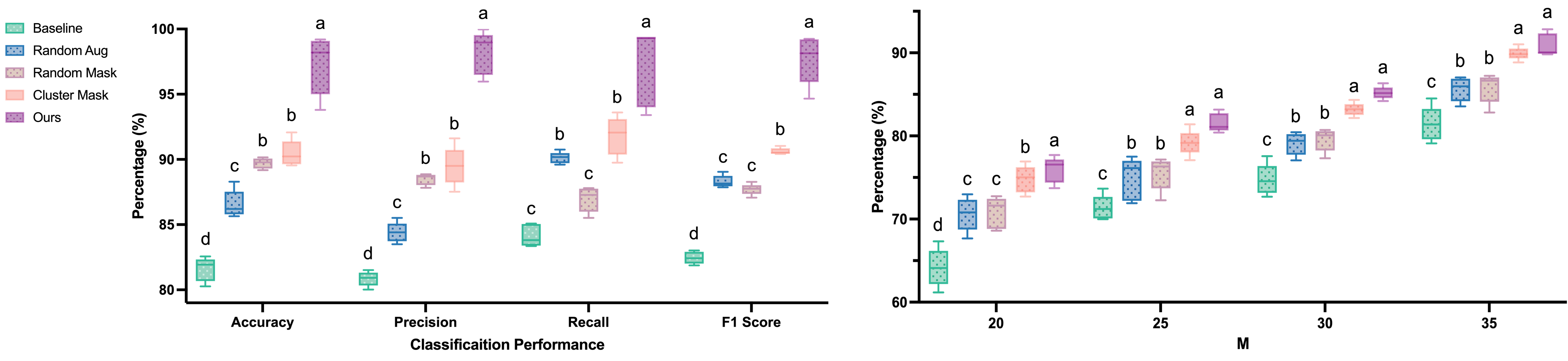}}
\caption{Tukey HSD post-hoc test results for MSL dataset using FCN. The left and right plots illustrate classification and attention accuracy, respectively. Methods not sharing any letter are significantly different by the Tukey-test at the 5\% level of significance. Note: y-axes are truncated to emphasize fine-grained differences.}
\label{fig:Obj1}
\end{figure*}

\begin{figure*}[t!]
\centerline{\includegraphics[width=1\textwidth]{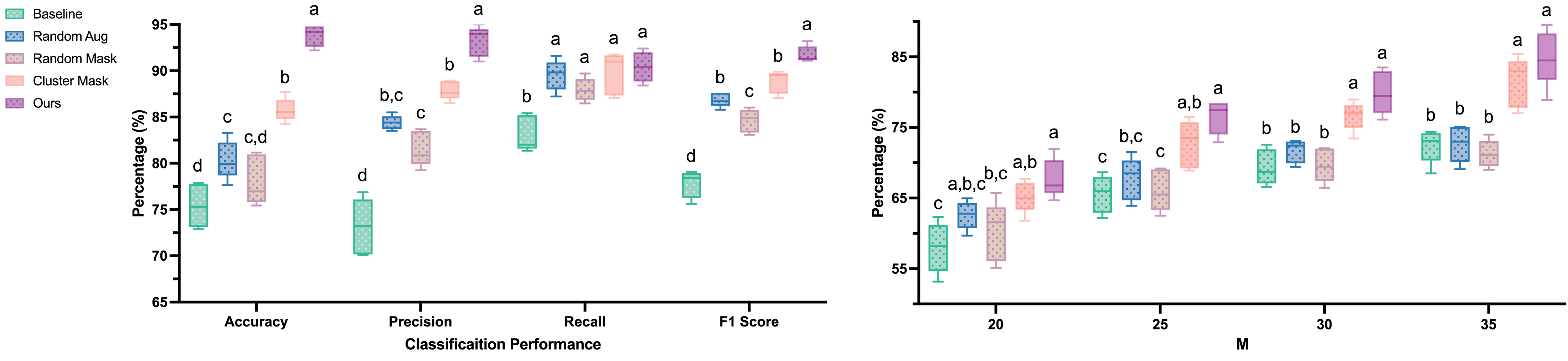}}
\caption{Tukey HSD post-hoc test results for STEVE dataset using FCN. 
The left and right plots illustrate classification and attention accuracy, respectively. Methods not sharing any letter are significantly different by the Tukey-test at the 5\% level of significance. Note: y-axes are truncated to emphasize fine-grained differences.}
\label{fig:Obj2}
\end{figure*}

% \begin{table}[htbp]
% \caption{Results for the T-test of Objective Permformance}
% \label{tab:T for Likert}
% \resizebox{\columnwidth}{!}{%
% \begin{tabular}{cccc}
% \toprule[0.8pt]
% Pairwise     & Automatic vs Cluster Mask & Automatic vs Ours & Cluster Mask vs Ours \\ \hline
% Intuition    &   $<$0.01                        &   $<$0.01                &  1                    \\
% Transparency &   $<$0.01                        &  $<$0.01                 &   1                   \\
% Reliability  &   $<$0.01                        &  $<$0.01                 &     $<$0.01                 \\ \bottomrule[0.8pt]
% \end{tabular}%
% }
% \end{table}

\subsection{Quantitative evaluation}

In order to effectively measure the objective performance difference between purely automatic algorithm-driven and human-machine cooperative anomaly detection methods, we conducted an analysis using Analysis of Variance (ANOVA) tests~\cite{st1989analysis}. 
% This methodological choice is crucial for examining the incremental benefits conferred by human intervention in algorithm-driven decision-making processes. 
Objective performance is divided into two aspects, namely classification correctness and attention correctness. For the former, we utilize the ANOVA test to compare the means of accuracy, precision, recall, and F1 score across different models to statistically ascertain any significant performance variations. 
For the latter, we employed the ANOVA test to determine whether there are statistically significant differences in relevance accuracy at varying levels of $M$, specifically at $M=20, 25, 30$, and $35$.
Upon identifying significant differences with the ANOVA test, we proceeded with the Tukey HSD post-hoc test to further distinguish which specific group means, among the compared groups, are significantly different from each other, ensuring a thorough evaluation of pairwise contrasts.

\begin{figure*}[t!]
\centerline{\includegraphics[width=1\textwidth]{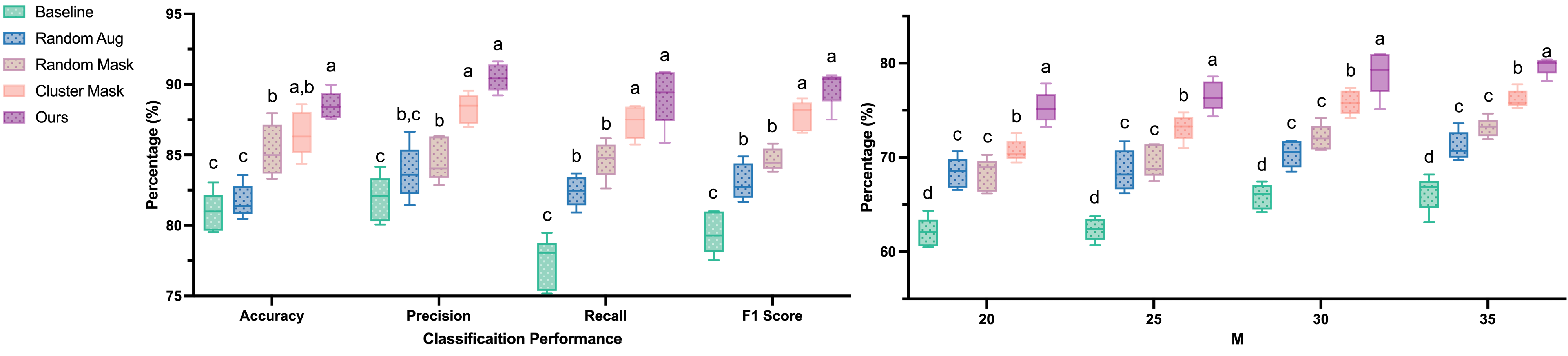}}
\caption{Tukey HSD post-hoc test results for KPI dataset using FCN. The left and right plots illustrate classification and attention accuracy, respectively. Methods not sharing any letter are significantly different by the Tukey-test at the 5\% level of significance. Note: y-axes are truncated to emphasize fine-grained differences.}
\label{fig:Obj3}
\end{figure*}

\begin{figure*}[t!]
\centerline{\includegraphics[width=1\textwidth]{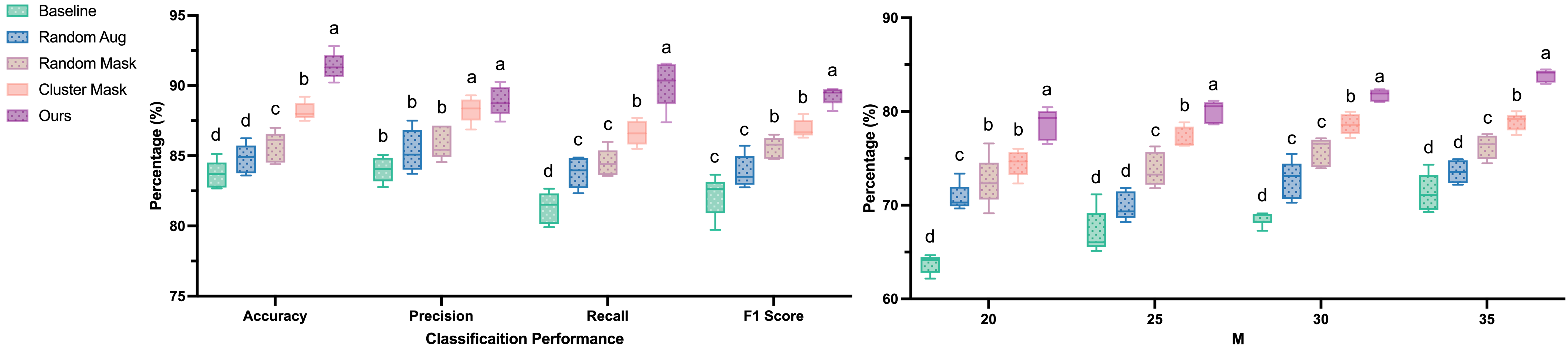}}
\caption{Tukey HSD post-hoc test results for KPI dataset using TST. The left and right plots illustrate classification and attention accuracy, respectively. Methods not sharing any letter are significantly different by the Tukey-test at the 5\% level of significance. Note: y-axes are truncated to emphasize fine-grained differences.}
\label{fig:Obj4}
\end{figure*}

We report the Tukey HSD post-hoc test results on the MSL and STEVE datasets using FCN in Figs~\ref{fig:Obj1} and~\ref{fig:Obj2}. For the more challenging real-world KPI dataset, we include both FCN and TST models, with results shown in Figs.~\ref{fig:Obj3} and~\ref{fig:Obj4}. Across all settings, \systemname{}~consistently outperforms baseline and augmentation-based methods in both prediction accuracy and attribution alignment. These results underscore the importance of evaluating not only classification outcomes but also the underlying rationale, reinforcing the need for human-in-the-loop validation.

\subsection{Annotation efficiency evaluation}

The results are reported in Table~\ref{tab:annotime}, which includes the average annotation time, annotated data size, the spuriousness label coverage (how many instances have been assigned a label), and the F1 gain in percentage compared to the basic model. We can observe that \systemname{}~consistently achieved the best trade-off: requiring the least annotation time (8.3 $\pm$ 2.1 min), while delivering the highest F1 improvement (18.485 $\pm$ 2.7\%) and covering over 800 sequences thanks to our propagation design. In contrast, despite much larger annotation budgets being used, the active learning baselines still underperformed because they solely rely on manual annotations, leading to significantly less spuriousness label coverage.
These results demonstrate that our framework supports minimal but essential human input. With fewer human annotations, our design of attribution-guided behavior summarization and annotation propagation leverages human efforts at a lower cost, which is not achievable through unstructured or conventional data sampling.

\begin{table}[t!]
\centering
\resizebox{\columnwidth}{!}{%
\begin{tabular}{ccccc}
\toprule[0.8pt]
Method       & Avg. annotation time (minutes) & Avg. annotation number & Coverage & F1 Gain (\%)    \\ \hline
random       & 14.2 $\pm$ 1.8                    & 160                    & 160      & 1.908 $\pm$ 1.2  \\
uncertainty  & 13.3 $\pm$ 1.4                    & 160                    & 160      & 2.610 $\pm$ 0.4  \\
diversity    & 14.7 $\pm$ 2.3                    & 160                    & 160      & 2.328 $\pm$ 1.9  \\
\systemname{}~(ours) & 8.3 $\pm$ 2.1                     & 10.2 $\pm$ 3.4            & 800      & 18.485 $\pm$ 2.7 \\ \bottomrule[0.8pt]
\end{tabular}%
}
\caption{Comparison of annotation time, number, coverage, and resulting F1 gain across \systemname{}~and baseline annotation strategies on MSL dataset.}
\label{tab:annotime}
\end{table}

\subsection{Objective evaluation}

% Fig.~\ref{fig: Likert} summarizes participants’ subjective evaluations of three kinds of spuriousness mitigation methods, Automatic Methods, Cluster Mask, and \systemname{} based on a five-point Likert scale. ``Automatic Methods'' refers to The questions are grouped into three categories: (1) intuitiveness of the method, (2) transparency of its process, and (3) perceived reliability of both process and outcome.

\begin{figure}[t!]
\centerline{\includegraphics[width=0.5\columnwidth]{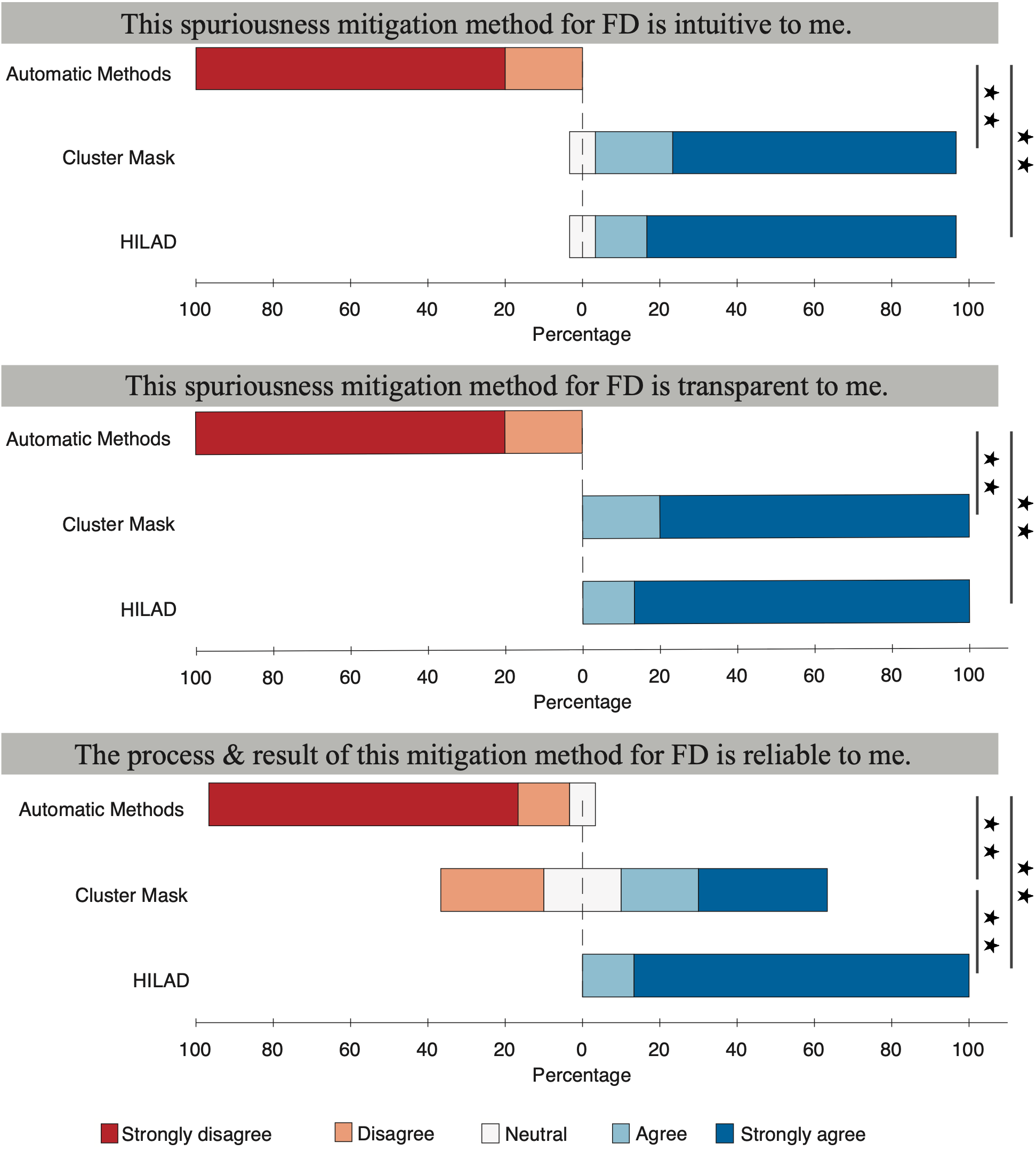}}
\caption{Participants perception of the system, according to Likert-Type Questions using a 5-point scale.}
\label{fig: Likert}
\end{figure}

The result of the five-point Likert scale for objective evaluation is shown in Fig.~\ref{fig: Likert}.
For a more comprehensive analysis, we also furnish the post-test results for the three methods, assessing their comparative effectiveness using a t-test.
The stars intended to flag levels of significance, which are based on the adjusted p-values.

\subsection{Ablation study}

\begin{figure}[t!]
\centerline{\includegraphics[width=\columnwidth]{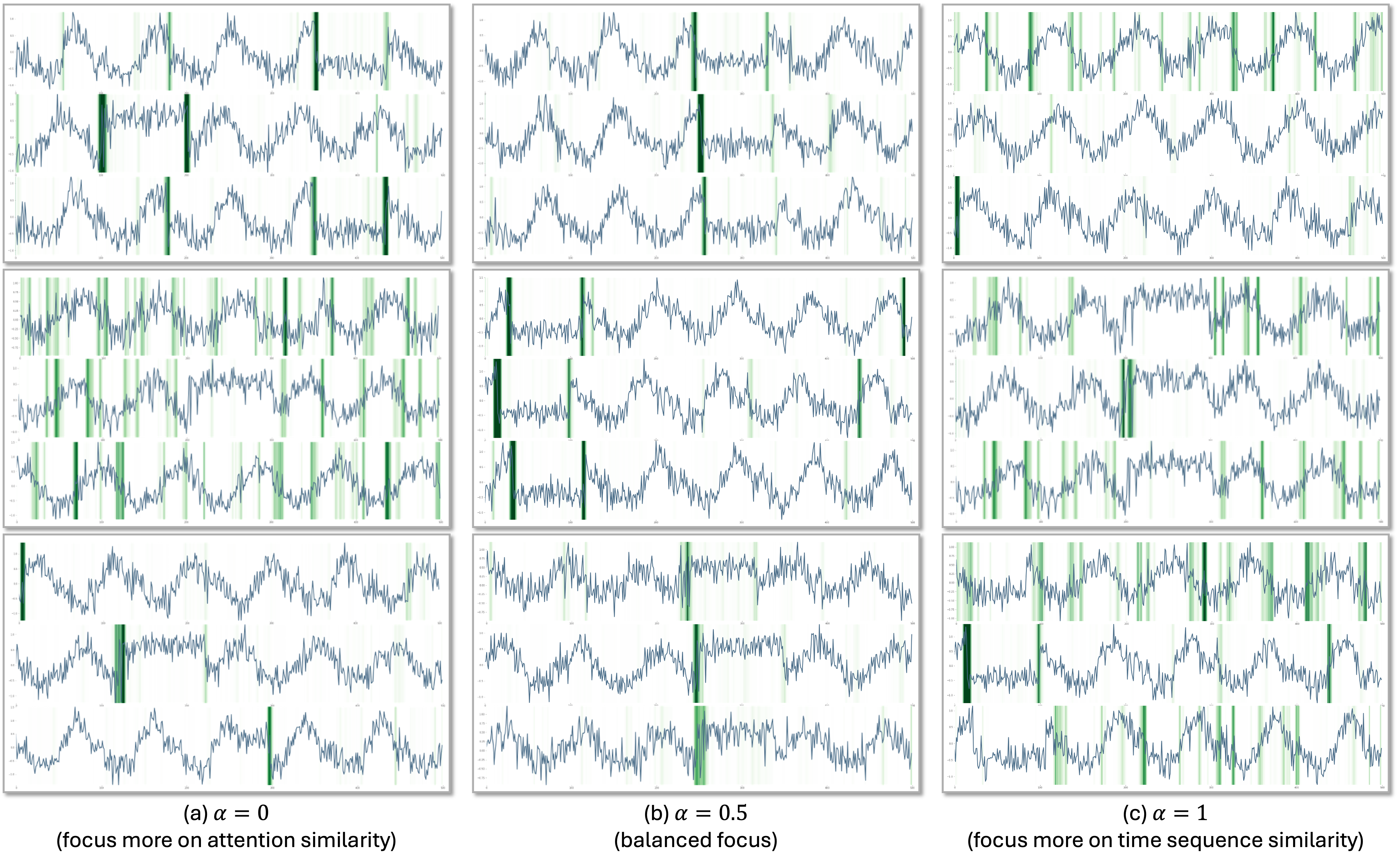}}
\caption{Qualitative observation on randomly sampled clusters when using different $\alpha$ values. We can observe that $\alpha=0$ leads to clusters with more coherent model attention only, while $\alpha=1$ produces clusters with consistent time sequences but dissimilar model behavior patterns. In contrast, our choice of $\alpha=0.5$ provides clusters with both consistent model attribution and similar time sequences.}
\label{fig:ablation_alpha}
\end{figure}

To validate our design choice of $\alpha$ = 0.5 in the joint similarity computation for behavior clustering, we conducted a sensitivity analysis by varying $\alpha$ across a range of values. Refer to Formula~\ref{eqn:dist_alpha} in Sec.~\ref{subsec:atttribution_sim}, $\alpha$ controls the balance between time sequence similarity (via DTW) and attribution similarity (via CAM), with higher $\alpha$ favoring time sequence similarity.
We first vary $\alpha$ across $\{0.0, 0.5, 1.0\}$ and provide qualitative observations on the produced clusters in Fig.~\ref{fig:ablation_alpha}. Specifically, in each setting, we randomly sample three clusters from the results and observe their included instances. The results show $\alpha=0.0$ overlooks the data consistency and $\alpha=1.0$ contrarily misses the model attribution consistency, while our choice $\alpha=0.5$ produces clusters with internally consistent time sequences and model attributes simultaneously.

We also quantitatively evaluate the impact of $\alpha$ on model performance, where we utilize FCN as the baseline on MSL data, measuring classification performance and attention correctness. In the experiment, we only allow cluster-level annotation and disable the adjustment of individual instances to ensure valid evaluation of $\alpha$. As shown in Table~\ref{tab:alpha}, $\alpha=0.5$ consistently produces the highest F1 and attention scores, supporting the use of equal weighting between sequence context and model behavior. Extreme values (e.g., $\alpha=0.0$ or $1.0$) led to performance drops, where $\alpha=1.0$ (only consider data similarity) results in the worst performance, as we lost neighbor consistency regarding model attributions. This experiment validates our parameter choice, confirming that collectively considering both temporal patterns and model attention is critical for meaningful behavior grouping.

\section{Discussion}
\subsection{Hypothesis 1}

Firstly, regarding classification correctness, Fig.~\ref{fig:Obj1} demonstrates that on univariate data, all data augmentation methods significantly enhance the model's classification performance compared to the baseline model.
Among these methods, within the context of human-AI cooperation, the \textbf{Cluster Mask} approach does not exhibit a significant difference from the \textbf{Random Mask} in the automatic method across other metrics, with the exception of the F1 score.
However, our method demonstrated significant enhancements in all four metrics of classification performance when compared to all other methods.
Compared to approach \textbf{Cluster Mask}, the improvement in our approach is facilitated by the ~\mybox{\textcolor{white}{D}} window's functionality, which enables humans to further refine the automatic clustering algorithm. This aspect also contributes to the observed large variance in our method across the four metrics, reflecting the variability in human judgment and adjustments. This variability is partly due to the high number of instances in each cluster. Despite this, in terms of overall performance, the adjustments made by humans do yield beneficial outcomes.

As depicted in Fig.~\ref{fig:Obj2}, similar to the univariate data, all data augmentation methods, with the exception of the accuracy metric for the \textbf{Random Mask} approach, show results that are significantly different from the baseline model.
Furthermore, within the realm of human-AI cooperation, the \textbf{Cluster Mask} approach exhibits a significant difference over other automatic algorithms solely in terms of accuracy.
However, our method demonstrates significant differences from both other automatic data augmentation methods and the \textbf{Cluster Mask} method in terms of accuracy, precision, and F1 score.
In terms of recall, no significant differences are observed among all data augmentation methods. This phenomenon can be attributed to the imbalanced nature of the dataset, where the abundance of normal instances prompts the model to occasionally misclassify abnormal instances as normal. In such imbalanced datasets, neither recall nor precision in isolation serves as a reliable measure of method performance. Instead, the F1 score emerges as a more accurate metric due to its ability to strike a balance between precision and recall.

Shifting our focus to attention accuracy, both the cluster mask and our method exhibit significant deviations from other automatic algorithms when applied to univariate data. However, as the data dimensionality increases, achieving accurate attention becomes increasingly challenging, particularly at $M=20$, where no significant disparities exist between the cluster mask method and other automatic algorithms. Nevertheless, consistent disparities persist between our methods and the automatic algorithm.
A more in-depth analysis reveals that, despite the notable improvements in classification performance when compared to the baseline model, the automatic algorithm fails to significantly enhance attention accuracy, especially concerning multivariate data. This observation reinforces the notion that our method not only objectively enhances classification performance but also elevates attention accuracy, thereby bolstering overall system reliability.

\begin{table}[t!]
\centering
\resizebox{\columnwidth}{!}{%
\begin{tabular}{p{1cm}<{\centering}p{2.3cm}<{\centering}p{2.3cm}<{\centering}p{2.3cm}<{\centering}p{2.3cm}<{\centering}p{2.3cm}<{\centering}}
\toprule[0.8pt]
    & $\alpha$=0.00 & $\alpha$=0.25 & \textbf{$\alpha$=0.50} & $\alpha$=0.75 & $\alpha$=1.0 \\ \hline
F1  & 88.21 $\pm$ 0.26              & 89.92 $\pm$ 0.27              & 90.62 $\pm$ 0.25                         & 88.33 $\pm$ 0.60              & 87.1 $\pm$ 0.76              \\
ATT & 86.43 $\pm$ 0.79              & 89.13 $\pm$ 0.78              & 89.90 $\pm$ 0.77                         & 85.82 $\pm$ 0.85              & 85.66 $\pm$ 0.84             \\ \bottomrule[0.8pt]
\end{tabular}%
}
\caption{Quantitative evaluation on the effect of the $\alpha$ parameter that controls the balance between temporal and attribution similarity in clustering, showing $\alpha=0.5$ achieves the best trade-off in both classification (F1) and attention (ATT) performance.}
\label{tab:alpha}
\end{table}

In summary, our method exhibits substantial deviations in both classification correctness and attention correctness when compared to the automatic algorithm across the two datasets. This substantiates our conjecture and underscores the empirical evidence supporting our approach.

\subsection{Hypothesis 2}

In the subjective evaluation, both the cluster mask method and our method are significantly different from the automatic algorithm in intuition, transparency, and reliability.
Because the automatic algorithm only provides classification results, it neither reveals the underlying decision criteria nor supports interaction.
The cluster mask method is not significantly different from intuition and transparency, as both provide reasons for the model to make a prediction.
But on reliability, we are significantly different from the cluster mask method because of the error of the automatic clustering algorithm itself, while in our method, the \mybox{\textcolor{white}{D}} window can provide more detailed options to further correct the automatic algorithm, which helps improve performance objectively. At the same time, humans are subjectively more likely to believe that our methods are more reliable.

Therefore, we verified that subjectively human-AI method has significant improvement in intuition, transparency, and reliability compared with automatic algorithms, thus verifying Hypothesis 2. Further, on the objective and subjective level, the necessity of introducing \mybox{\textcolor{white}{D}} window is verified by comparing with \textbf{Cluster Mask} method.

\section{Conclusion}
In conclusion, we introduce a human-AI collaboration framework, \systemname{}, to enhance the performance and reliability of time series anomaly detection models. With our interactive visual interface, human expertise can be effectively leveraged to detect and eliminate the hidden model biases in scale. Through objective and subjective evaluations with two benchmark datasets, we demonstrate the exceptional capabilities of \systemname{} in enhancing the model's classification accuracy and attention correctness. With \systemname{}, we underscore the potential of human-AI collaboration in supporting greater transparency, reliability, and trustworthiness of machine learning models. We hope the design of how human expertise can be effectively leveraged in our work can inspire the development of next-generation trustworthy AI systems.

\bibliographystyle{unsrtnat}
\small
\bibliography{references}

\begin{thebibliography}{69}
\providecommand{\natexlab}[1]{#1}
\providecommand{\url}[1]{\texttt{#1}}
\expandafter\ifx\csname urlstyle\endcsname\relax
  \providecommand{\doi}[1]{doi: #1}\else
  \providecommand{\doi}{doi: \begingroup \urlstyle{rm}\Url}\fi

\bibitem[Ahmed et~al.(2016)Ahmed, Mahmood, and Islam]{ahmed2016survey}
Mohiuddin Ahmed, Abdun~Naser Mahmood, and Md~Rafiqul Islam.
\newblock A survey of anomaly detection techniques in financial domain.
\newblock \emph{Future Generation Computer Systems}, 55:\penalty0 278--288, 2016.

\bibitem[{\v{S}}abi{\'c} et~al.(2021){\v{S}}abi{\'c}, Keeley, Henderson, and Nannemann]{vsabic2021healthcare}
Edin {\v{S}}abi{\'c}, David Keeley, Bailey Henderson, and Sara Nannemann.
\newblock Healthcare and anomaly detection: using machine learning to predict anomalies in heart rate data.
\newblock \emph{AI \& SOCIETY}, 36\penalty0 (1):\penalty0 149--158, 2021.

\bibitem[Li et~al.(2019)Li, Chen, Jin, Shi, Goh, and Ng]{li2019mad}
Dan Li, Dacheng Chen, Baihong Jin, Lei Shi, Jonathan Goh, and See-Kiong Ng.
\newblock Mad-gan: Multivariate anomaly detection for time series data with generative adversarial networks.
\newblock In \emph{International conference on artificial neural networks}, pages 703--716. Springer, 2019.

\bibitem[Dwivedi et~al.(2023)Dwivedi, Dave, Naik, Singhal, Omer, Patel, Qian, Wen, Shah, Morgan, et~al.]{dwivedi2023explainable}
Rudresh Dwivedi, Devam Dave, Het Naik, Smiti Singhal, Rana Omer, Pankesh Patel, Bin Qian, Zhenyu Wen, Tejal Shah, Graham Morgan, et~al.
\newblock Explainable ai (xai): Core ideas, techniques, and solutions.
\newblock \emph{ACM Computing Surveys}, 55\penalty0 (9):\penalty0 1--33, 2023.

\bibitem[Xuan et~al.(2024{\natexlab{a}})Xuan, Deng, Lin, Kong, and Ma]{xuan2024suny}
Xiwei Xuan, Ziquan Deng, Hsuan-Tien Lin, Zhaodan Kong, and Kwan-Liu Ma.
\newblock Suny: A visual interpretation framework for convolutional neural networks from a necessary and sufficient perspective.
\newblock In \emph{Proceedings of the IEEE/CVF Conference on Computer Vision and Pattern Recognition}, pages 8371--8376, 2024{\natexlab{a}}.

\bibitem[Zhou et~al.(2016)Zhou, Khosla, Lapedriza, Oliva, and Torralba]{zhou2016learning}
Bolei Zhou, Aditya Khosla, Agata Lapedriza, Aude Oliva, and Antonio Torralba.
\newblock Learning deep features for discriminative localization.
\newblock In \emph{Proceedings of the IEEE conference on computer vision and pattern recognition}, pages 2921--2929, 2016.

\bibitem[Boniol et~al.(2023)Boniol, Meftah, Remy, Didier, and Palpanas]{boniol2023dcnn}
Paul Boniol, Mohammed Meftah, Emmanuel Remy, Bruno Didier, and Themis Palpanas.
\newblock dcnn/dcam: anomaly precursors discovery in multivariate time series with deep convolutional neural networks.
\newblock \emph{Data-Centric Engineering}, 4:\penalty0 e30, 2023.

\bibitem[Barkouki et~al.(2023)Barkouki, Deng, Karasinski, Kong, and Robinson]{barkouki2023xai}
Tammer Barkouki, Ziquan Deng, John Karasinski, Zhaodan Kong, and Stephen Robinson.
\newblock Xai design goals and evaluation metrics for space exploration: A survey of human spaceflight domain experts.
\newblock In \emph{AIAA SCITECH 2023 Forum}, page 1828, 2023.

\bibitem[Amershi et~al.(2019)Amershi, Weld, Vorvoreanu, Fourney, Nushi, Collisson, Suh, Iqbal, Bennett, Inkpen, et~al.]{amershi2019guidelines}
Saleema Amershi, Dan Weld, Mihaela Vorvoreanu, Adam Fourney, Besmira Nushi, Penny Collisson, Jina Suh, Shamsi Iqbal, Paul~N Bennett, Kori Inkpen, et~al.
\newblock Guidelines for human-ai interaction.
\newblock In \emph{Proceedings of the 2019 chi conference on human factors in computing systems}, pages 1--13, 2019.

\bibitem[Xuan et~al.(2025{\natexlab{a}})Xuan, Wang, He, Ono, Gou, Ma, and Ren]{xuan2025vista}
Xiwei Xuan, Xiaoqi Wang, Wenbin He, Jorge~Piazentin Ono, Liang Gou, Kwan-Liu Ma, and Liu Ren.
\newblock Vista: A visual analytics framework to enhance foundation model-generated data labels.
\newblock \emph{IEEE Transactions on Visualization and Computer Graphics}, 2025{\natexlab{a}}.

\bibitem[Xuan et~al.(2025{\natexlab{b}})Xuan, Ono, Gou, Ma, and Ren]{xuan2024attributionscanner}
Xiwei Xuan, Jorge~Piazentin Ono, Liang Gou, Kwan-Liu Ma, and Liu Ren.
\newblock Attributionscanner: A visual analytics system for model validation with metadata-free slice finding.
\newblock \emph{IEEE Transactions on Visualization and Computer Graphics}, 2025{\natexlab{b}}.

\bibitem[Kieu et~al.(2019)Kieu, Yang, Guo, and Jensen]{kieu2019outlier}
Tung Kieu, Bin Yang, Chenjuan Guo, and Christian~S Jensen.
\newblock Outlier detection for time series with recurrent autoencoder ensembles.
\newblock In \emph{IJCAI}, pages 2725--2732, 2019.

\bibitem[Malhotra et~al.(2015)Malhotra, Vig, Shroff, Agarwal, et~al.]{malhotra2015long}
Pankaj Malhotra, Lovekesh Vig, Gautam Shroff, Puneet Agarwal, et~al.
\newblock Long short term memory networks for anomaly detection in time series.
\newblock In \emph{Esann}, volume 2015, page~89, 2015.

\bibitem[Taylor and Letham(2018)]{taylor2018forecasting}
Sean~J Taylor and Benjamin Letham.
\newblock Forecasting at scale.
\newblock \emph{The American Statistician}, 72\penalty0 (1):\penalty0 37--45, 2018.

\bibitem[Tuli et~al.(2022)Tuli, Casale, and Jennings]{tuli2022tranad}
Shreshth Tuli, Giuliano Casale, and Nicholas~R. Jennings.
\newblock Tranad: deep transformer networks for anomaly detection in multivariate time series data.
\newblock \emph{Proc. VLDB Endow.}, 15\penalty0 (6):\penalty0 1201–1214, February 2022.
\newblock ISSN 2150-8097.

\bibitem[Del~Campo et~al.(2021)Del~Campo, Neri, Villegas, S{\'a}nchez, Dom{\'\i}nguez, and Jim{\'e}nez]{del2021auto}
Felipe~Arias Del~Campo, Mar{\'\i}a Cristina~Guevara Neri, Osslan Osiris~Vergara Villegas, Vianey Guadalupe~Cruz S{\'a}nchez, Humberto de Jes{\'u}s~Ochoa Dom{\'\i}nguez, and Vicente~Garc{\'\i}a Jim{\'e}nez.
\newblock Auto-adaptive multilayer perceptron for univariate time series classification.
\newblock \emph{Expert Systems with Applications}, 181:\penalty0 115147, 2021.

\bibitem[Gallicchio et~al.(2017)Gallicchio, Micheli, Silvestri, et~al.]{gallicchio2017local}
Claudio Gallicchio, Alessio Micheli, Luca Silvestri, et~al.
\newblock Local lyapunov exponents of deep rnn.
\newblock In \emph{Proceedings of the 25th European Symposium on Artificial Neural Networks (ESANN)}, pages 559--564. i6doc. com publication, 2017.

\bibitem[Zhao et~al.(2017)Zhao, Lu, Chen, Liu, and Wu]{zhao2017convolutional}
Bendong Zhao, Huanzhang Lu, Shangfeng Chen, Junliang Liu, and Dongya Wu.
\newblock Convolutional neural networks for time series classification.
\newblock \emph{Journal of Systems Engineering and Electronics}, 28\penalty0 (1):\penalty0 162--169, 2017.

\bibitem[Nabrawi and Alanazi(2023)]{nabrawi2023fraud}
Eman Nabrawi and Abdullah Alanazi.
\newblock Fraud detection in healthcare insurance claims using machine learning.
\newblock \emph{Risks}, 11\penalty0 (9):\penalty0 160, 2023.

\bibitem[Meena and Roy(2022)]{meena2022bone}
Tanushree Meena and Sudipta Roy.
\newblock Bone fracture detection using deep supervised learning from radiological images: A paradigm shift.
\newblock \emph{Diagnostics}, 12\penalty0 (10):\penalty0 2420, 2022.

\bibitem[Uddin et~al.(2019)Uddin, Khan, Hossain, and Moni]{uddin2019comparing}
Shahadat Uddin, Arif Khan, Md~Ekramul Hossain, and Mohammad~Ali Moni.
\newblock Comparing different supervised machine learning algorithms for disease prediction.
\newblock \emph{BMC medical informatics and decision making}, 19\penalty0 (1):\penalty0 1--16, 2019.

\bibitem[Dosovitskiy et~al.(2020)Dosovitskiy, Beyer, Kolesnikov, Weissenborn, Zhai, Unterthiner, Dehghani, Minderer, Heigold, Gelly, et~al.]{dosovitskiy2020image}
Alexey Dosovitskiy, Lucas Beyer, Alexander Kolesnikov, Dirk Weissenborn, Xiaohua Zhai, Thomas Unterthiner, Mostafa Dehghani, Matthias Minderer, Georg Heigold, Sylvain Gelly, et~al.
\newblock An image is worth 16x16 words: Transformers for image recognition at scale.
\newblock \emph{arXiv preprint arXiv:2010.11929}, 2020.

\bibitem[Liu et~al.(2021{\natexlab{a}})Liu, Lin, Cao, Hu, Wei, Zhang, Lin, and Guo]{liu2021swin}
Ze~Liu, Yutong Lin, Yue Cao, Han Hu, Yixuan Wei, Zheng Zhang, Stephen Lin, and Baining Guo.
\newblock Swin transformer: Hierarchical vision transformer using shifted windows.
\newblock In \emph{Proceedings of the IEEE/CVF international conference on computer vision}, pages 10012--10022, 2021{\natexlab{a}}.

\bibitem[Liu et~al.(2021{\natexlab{b}})Liu, Ren, Ma, Jiao, Chen, Wang, and Song]{liu2021gated}
Minghao Liu, Shengqi Ren, Siyuan Ma, Jiahui Jiao, Yizhou Chen, Zhiguang Wang, and Wei Song.
\newblock Gated transformer networks for multivariate time series classification.
\newblock \emph{arXiv preprint arXiv:2103.14438}, 2021{\natexlab{b}}.

\bibitem[Yang et~al.(2021)Yang, Tsai, and Chen]{yang2021voice2series}
Chao-Han~Huck Yang, Yun-Yun Tsai, and Pin-Yu Chen.
\newblock Voice2series: Reprogramming acoustic models for time series classification.
\newblock In \emph{International conference on machine learning}, pages 11808--11819. PMLR, 2021.

\bibitem[Jung et~al.(2021)Jung, Ramanan, Amjadi, Karingula, Taylor, and Coelho~Jr]{jung2021time}
Deokwoo Jung, Nandini Ramanan, Mehrnaz Amjadi, Sankeerth~Rao Karingula, Jake Taylor, and Claudionor~Nunes Coelho~Jr.
\newblock Time series anomaly detection with label-free model selection.
\newblock \emph{arXiv preprint arXiv:2106.07473}, 2021.

\bibitem[Wang et~al.(2024)Wang, He, Xuan, Sebastian, Ono, Li, Behpour, Doan, Gou, Shen, et~al.]{wang2024use}
Xiaoqi Wang, Wenbin He, Xiwei Xuan, Clint Sebastian, Jorge~Piazentin Ono, Xin Li, Sima Behpour, Thang Doan, Liang Gou, Han-Wei Shen, et~al.
\newblock Use: Universal segment embeddings for open-vocabulary image segmentation.
\newblock In \emph{Proceedings of the IEEE/CVF Conference on Computer Vision and Pattern Recognition}, pages 4187--4196, 2024.

\bibitem[Goswami et~al.(2022)Goswami, Challu, Callot, Minorics, and Kan]{goswami2022unsupervised}
Mononito Goswami, Cristian Challu, Laurent Callot, Lenon Minorics, and Andrey Kan.
\newblock Unsupervised model selection for time-series anomaly detection.
\newblock \emph{arXiv preprint arXiv:2210.01078}, 2022.

\bibitem[Deng and Kong(2021)]{deng2021interpretable}
Ziquan Deng and Zhaodan Kong.
\newblock Interpretable fault diagnosis for cyberphysical systems: A learning perspective.
\newblock \emph{Computer}, 54\penalty0 (9):\penalty0 30--38, 2021.

\bibitem[Xuan et~al.(2024{\natexlab{b}})Xuan, Deng, Lin, and Ma]{xuan2024slim}
Xiwei Xuan, Ziquan Deng, Hsuan-Tien Lin, and Kwan-Liu Ma.
\newblock Slim: Spuriousness mitigation with minimal human annotations.
\newblock In \emph{European Conference on Computer Vision}, pages 215--231. Springer, 2024{\natexlab{b}}.

\bibitem[Schaekermann et~al.(2020)Schaekermann, Cai, Huang, and Sayres]{schaekermann2020expert}
Mike Schaekermann, Carrie~J Cai, Abigail~E Huang, and Rory Sayres.
\newblock Expert discussions improve comprehension of difficult cases in medical image assessment.
\newblock In \emph{Proceedings of the 2020 CHI conference on human factors in computing systems}, pages 1--13, 2020.

\bibitem[Xuan et~al.(2022)Xuan, Zhang, Kwon, and Ma]{xuan2022vac}
Xiwei Xuan, Xiaoyu Zhang, Oh-Hyun Kwon, and Kwan-Liu Ma.
\newblock Vac-cnn: A visual analytics system for comparative studies of deep convolutional neural networks.
\newblock \emph{IEEE Transactions on Visualization and Computer Graphics}, 28\penalty0 (6):\penalty0 2326--2337, 2022.

\bibitem[Zhang et~al.(2023)Zhang, Xuan, Dima, Sexton, and Ma]{zhang2023labelvizier}
Xiaoyu Zhang, Xiwei Xuan, Alden Dima, Thurston Sexton, and Kwan-Liu Ma.
\newblock Labelvizier: Interactive validation and relabeling for technical text annotations.
\newblock In \emph{2023 IEEE 16th Pacific Visualization Symposium (PacificVis)}, pages 167--176. IEEE, 2023.

\bibitem[Yildirim et~al.(2024)Yildirim, Richardson, Wetscherek, Bajwa, Jacob, Pinnock, Harris, Coelho De~Castro, Bannur, Hyland, et~al.]{yildirim2024multimodal}
Nur Yildirim, Hannah Richardson, Maria~Teodora Wetscherek, Junaid Bajwa, Joseph Jacob, Mark~Ames Pinnock, Stephen Harris, Daniel Coelho De~Castro, Shruthi Bannur, Stephanie Hyland, et~al.
\newblock Multimodal healthcare ai: identifying and designing clinically relevant vision-language applications for radiology.
\newblock In \emph{Proceedings of the CHI Conference on Human Factors in Computing Systems}, pages 1--22, 2024.

\bibitem[Deng et~al.(2025)Deng, Prugsanapan, and Kong]{deng2025anytime}
Ziquan Deng, Pinn Prugsanapan, and Zhaodan Kong.
\newblock Anytime communication: A human-ai collaboration framework for causal-based root cause analysis.
\newblock In \emph{AIAA SCITECH 2025 Forum}, page 2252, 2025.

\bibitem[Yang et~al.(2022)Yang, Ye, Zhang, Xiao, Xia, Wang, Zhu, Pfister, and Liu]{yang2022diagnosing}
Weikai Yang, Xi~Ye, Xingxing Zhang, Lanxi Xiao, Jiazhi Xia, Zhongyuan Wang, Jun Zhu, Hanspeter Pfister, and Shixia Liu.
\newblock Diagnosing ensemble few-shot classifiers.
\newblock \emph{IEEE Transactions on Visualization and Computer Graphics}, 28\penalty0 (9):\penalty0 3292--3306, 2022.

\bibitem[Yang et~al.(2023)Yang, Guo, Wu, Wang, Guo, Li, and Liu]{yang2023interactive}
Weikai Yang, Yukai Guo, Jing Wu, Zheng Wang, Lan-Zhe Guo, Yu-Feng Li, and Shixia Liu.
\newblock Interactive reweighting for mitigating label quality issues.
\newblock \emph{IEEE Transactions on Visualization and Computer Graphics}, 2023.

\bibitem[Li et~al.(2022{\natexlab{a}})Li, Wang, Yang, Wu, Zhang, Liu, Sun, Zhang, and Liu]{li2022unified}
Zhen Li, Xiting Wang, Weikai Yang, Jing Wu, Zhengyan Zhang, Zhiyuan Liu, Maosong Sun, Hui Zhang, and Shixia Liu.
\newblock A unified understanding of deep nlp models for text classification.
\newblock \emph{IEEE Transactions on Visualization and Computer Graphics}, 28\penalty0 (12):\penalty0 4980--4994, 2022{\natexlab{a}}.

\bibitem[He et~al.(2021)He, Zou, Shekar, Gou, and Ren]{he2021can}
Wenbin He, Lincan Zou, Arvind~Kumar Shekar, Liang Gou, and Liu Ren.
\newblock Where can we help? a visual analytics approach to diagnosing and improving semantic segmentation of movable objects.
\newblock \emph{IEEE Transactions on Visualization and Computer Graphics}, 28\penalty0 (1):\penalty0 1040--1050, 2021.

\bibitem[Gou et~al.(2020)Gou, Zou, Li, Hofmann, Shekar, Wendt, and Ren]{gou2020vatld}
Liang Gou, Lincan Zou, Nanxiang Li, Michael Hofmann, Arvind~Kumar Shekar, Axel Wendt, and Liu Ren.
\newblock Vatld: A visual analytics system to assess, understand and improve traffic light detection.
\newblock \emph{IEEE transactions on visualization and computer graphics}, 27\penalty0 (2):\penalty0 261--271, 2020.

\bibitem[Liu et~al.(2012)Liu, Zhou, Pan, Song, Qian, Cai, and Lian]{liu2012tiara}
Shixia Liu, Michelle~X Zhou, Shimei Pan, Yangqiu Song, Weihong Qian, Weijia Cai, and Xiaoxiao Lian.
\newblock Tiara: Interactive, topic-based visual text summarization and analysis.
\newblock \emph{ACM Transactions on Intelligent Systems and Technology (TIST)}, 3\penalty0 (2):\penalty0 1--28, 2012.

\bibitem[Yan et~al.(2025)Yan, Xuan, Ono, Guo, Mohanty, Kumar, Gou, Wang, and Ren]{yan2025vislix}
Xinyuan Yan, Xiwei Xuan, Jorge~Piazentin Ono, Jiajing Guo, Vikram Mohanty, Shekar~Arvind Kumar, Liang Gou, Bei Wang, and Liu Ren.
\newblock Vislix: An xai framework for validating vision models with slice discovery and analysis.
\newblock In \emph{Computer Graphics Forum}, page e70125. Wiley Online Library, 2025.

\bibitem[Malik et~al.(2025)Malik, Xuan, and Ma]{malik2025towards}
Divyanshu Malik, Xiwei Xuan, and Kwan-Liu Ma.
\newblock Towards interactive 3d surgical scene reconstruction: An incremental training and monitoring framework.
\newblock In \emph{2025 IEEE 22nd International Symposium on Biomedical Imaging (ISBI)}, pages 1--4. IEEE, 2025.

\bibitem[Geiger et~al.(2020)Geiger, Liu, Alnegheimish, Cuesta-Infante, and Veeramachaneni]{geiger2020tadgan}
Alexander Geiger, Dongyu Liu, Sarah Alnegheimish, Alfredo Cuesta-Infante, and Kalyan Veeramachaneni.
\newblock Tadgan: Time series anomaly detection using generative adversarial networks.
\newblock In \emph{2020 ieee international conference on big data (big data)}, pages 33--43. IEEE, 2020.

\bibitem[Alnegheimish et~al.(2022)Alnegheimish, Liu, Sala, Berti-Equille, and Veeramachaneni]{alnegheimish2022sintel}
Sarah Alnegheimish, Dongyu Liu, Carles Sala, Laure Berti-Equille, and Kalyan Veeramachaneni.
\newblock Sintel: A machine learning framework to extract insights from signals.
\newblock In \emph{Proceedings of the 2022 International Conference on Management of Data}, pages 1855--1865, 2022.

\bibitem[Liu et~al.(2022)Liu, Alnegheimish, Zytek, and Veeramachaneni]{liu2022mtv}
Dongyu Liu, Sarah Alnegheimish, Alexandra Zytek, and Kalyan Veeramachaneni.
\newblock Mtv: Visual analytics for detecting, investigating, and annotating anomalies in multivariate time series.
\newblock \emph{Proceedings of the ACM on Human-Computer Interaction}, 6\penalty0 (CSCW1):\penalty0 1--30, 2022.

\bibitem[Feng et~al.(2023)Feng, Hu, Yang, Polk, Zhao, Liu, and Yang]{feng2023timepool}
Tinghao Feng, Yueqi Hu, Jing Yang, Tom Polk, Ye~Zhao, Shixia Liu, and Zhaocong Yang.
\newblock Timepool: Visually answer “which and when” questions on univariate time series.
\newblock In \emph{2023 IEEE Visualization and Visual Analytics (VIS)}, pages 201--205. IEEE, 2023.

\bibitem[Ruta et~al.(2019)Ruta, Sawada, McKeough, Behrisch, and Beyer]{ruta2019sax}
Nicholas Ruta, Naoko Sawada, Katy McKeough, Michael Behrisch, and Johanna Beyer.
\newblock Sax navigator: Time series exploration through hierarchical clustering.
\newblock In \emph{2019 IEEE Visualization Conference (VIS)}, pages 236--240. IEEE, 2019.

\bibitem[Guo et~al.(2021)Guo, Jin, Chen, Gotz, Zha, and Cao]{guo2021interpretable}
Shunan Guo, Zhuochen Jin, Qing Chen, David Gotz, Hongyuan Zha, and Nan Cao.
\newblock Interpretable anomaly detection in event sequences via sequence matching and visual comparison.
\newblock \emph{IEEE Transactions on Visualization and Computer Graphics}, 28\penalty0 (12):\penalty0 4531--4545, 2021.

\bibitem[Fujiwara et~al.(2020)Fujiwara, Sakamoto, Nonaka, Yamamoto, Ma, et~al.]{fujiwara2020visual}
Takanori Fujiwara, Naohisa Sakamoto, Jorji Nonaka, Keiji Yamamoto, Kwan-Liu Ma, et~al.
\newblock A visual analytics framework for reviewing multivariate time-series data with dimensionality reduction.
\newblock \emph{IEEE transactions on visualization and computer graphics}, 27\penalty0 (2):\penalty0 1601--1611, 2020.

\bibitem[Hundman et~al.(2018)Hundman, Constantinou, Laporte, Colwell, and Soderstrom]{hundman2018detecting}
Kyle Hundman, Valentino Constantinou, Christopher Laporte, Ian Colwell, and Tom Soderstrom.
\newblock Detecting spacecraft anomalies using lstms and nonparametric dynamic thresholding.
\newblock In \emph{Proceedings of the 24th ACM SIGKDD international conference on knowledge discovery \& data mining}, pages 387--395, 2018.

\bibitem[Selvaraju et~al.(2020)Selvaraju, Cogswell, Das, Vedantam, Parikh, and Batra]{selvaraju2020grad}
Ramprasaath~R Selvaraju, Michael Cogswell, Abhishek Das, Ramakrishna Vedantam, Devi Parikh, and Dhruv Batra.
\newblock Grad-cam: visual explanations from deep networks via gradient-based localization.
\newblock \emph{International journal of computer vision}, 128:\penalty0 336--359, 2020.

\bibitem[Wang et~al.(2020)Wang, Wang, Du, Yang, Zhang, Ding, Mardziel, and Hu]{wang2020score}
Haofan Wang, Zifan Wang, Mengnan Du, Fan Yang, Zijian Zhang, Sirui Ding, Piotr Mardziel, and Xia Hu.
\newblock Score-cam: Score-weighted visual explanations for convolutional neural networks.
\newblock In \emph{Proceedings of the IEEE/CVF conference on computer vision and pattern recognition workshops}, pages 24--25, 2020.

\bibitem[Petsiuk et~al.(2018)Petsiuk, Das, and Saenko]{petsiuk2018rise}
Vitali Petsiuk, Abir Das, and Kate Saenko.
\newblock Rise: Randomized input sampling for explanation of black-box models.
\newblock \emph{arXiv preprint arXiv:1806.07421}, 2018.

\bibitem[McInnes et~al.(2018)McInnes, Healy, and Melville]{mcinnes2018umap}
Leland McInnes, John Healy, and James Melville.
\newblock Umap: Uniform manifold approximation and projection for dimension reduction.
\newblock \emph{arXiv preprint arXiv:1802.03426}, 2018.

\bibitem[Krishna and Murty(1999)]{krishna1999genetic}
K~Krishna and M~Narasimha Murty.
\newblock Genetic k-means algorithm.
\newblock \emph{IEEE Transactions on Systems, Man, and Cybernetics, Part B (Cybernetics)}, 29\penalty0 (3):\penalty0 433--439, 1999.

\bibitem[Zhou et~al.(2003)Zhou, Bousquet, Lal, Weston, and Sch{\"o}lkopf]{zhou2003learning}
Dengyong Zhou, Olivier Bousquet, Thomas Lal, Jason Weston, and Bernhard Sch{\"o}lkopf.
\newblock Learning with local and global consistency.
\newblock \emph{Advances in neural information processing systems}, 16, 2003.

\bibitem[Singla et~al.(2022)Singla, Moayeri, and Feizi]{singla2022core}
Sahil Singla, Mazda Moayeri, and Soheil Feizi.
\newblock Core risk minimization using salient imagenet.
\newblock \emph{arXiv preprint arXiv:2203.15566}, 2022.

\bibitem[Tang and Zhang(2022)]{tang2022mtsmae}
Peiwang Tang and Xianchao Zhang.
\newblock Mtsmae: Masked autoencoders for multivariate time-series forecasting.
\newblock In \emph{2022 IEEE 34th International Conference on Tools with Artificial Intelligence (ICTAI)}, pages 982--989. IEEE, 2022.

\bibitem[Wu and Keogh(2021)]{wu2021current}
Renjie Wu and Eamonn~J Keogh.
\newblock Current time series anomaly detection benchmarks are flawed and are creating the illusion of progress.
\newblock \emph{IEEE transactions on knowledge and data engineering}, 35\penalty0 (3):\penalty0 2421--2429, 2021.

\bibitem[Deng et~al.(2023)Deng, Eshima, Nabity, and Kong]{deng2023causal}
Ziquan Deng, Samuel~P Eshima, James Nabity, and Zhaodan Kong.
\newblock Causal signal temporal logic for the environmental control and life support system’s fault analysis and explanation.
\newblock \emph{IEEE Access}, 11:\penalty0 26471--26482, 2023.

\bibitem[Li et~al.(2022{\natexlab{b}})Li, Zhao, Zhang, Sun, Chen, Wen, Ma, and Pei]{li2022constructing}
Zeyan Li, Nengwen Zhao, Shenglin Zhang, Yongqian Sun, Pengfei Chen, Xidao Wen, Minghua Ma, and Dan Pei.
\newblock Constructing large-scale real-world benchmark datasets for aiops.
\newblock \emph{arXiv preprint arXiv:2208.03938}, 2022{\natexlab{b}}.

\bibitem[Wang et~al.(2017)Wang, Yan, and Oates]{wang2017time}
Zhiguang Wang, Weizhong Yan, and Tim Oates.
\newblock Time series classification from scratch with deep neural networks: A strong baseline.
\newblock In \emph{2017 International joint conference on neural networks (IJCNN)}, pages 1578--1585. IEEE, 2017.

\bibitem[Zerveas et~al.(2021)Zerveas, Jayaraman, Patel, Bhamidipaty, and Eickhoff]{zerveas2021transformer}
George Zerveas, Srideepika Jayaraman, Dhaval Patel, Anuradha Bhamidipaty, and Carsten Eickhoff.
\newblock A transformer-based framework for multivariate time series representation learning.
\newblock In \emph{Proceedings of the 27th ACM SIGKDD conference on knowledge discovery \& data mining}, pages 2114--2124, 2021.

\bibitem[Wickstr{\o}m et~al.(2020)Wickstr{\o}m, Mikalsen, Kampffmeyer, Revhaug, and Jenssen]{wickstrom2020uncertainty}
Kristoffer Wickstr{\o}m, Karl~{\O}yvind Mikalsen, Michael Kampffmeyer, Arthur Revhaug, and Robert Jenssen.
\newblock Uncertainty-aware deep ensembles for reliable and explainable predictions of clinical time series.
\newblock \emph{IEEE Journal of Biomedical and Health Informatics}, 25\penalty0 (7):\penalty0 2435--2444, 2020.

\bibitem[De~Winter and Dodou(2010)]{de2010five}
Joost~CF De~Winter and Dimitra Dodou.
\newblock Five-point likert items: t test versus mann-whitney-wilcoxon.
\newblock \emph{Practical assessment, research \& evaluation}, 15\penalty0 (11):\penalty0 1--12, 2010.

\bibitem[St et~al.(1989)St, Wold, et~al.]{st1989analysis}
Lars St, Svante Wold, et~al.
\newblock Analysis of variance (anova).
\newblock \emph{Chemometrics and intelligent laboratory systems}, 6\penalty0 (4):\penalty0 259--272, 1989.

\bibitem[Laptev et~al.(2015)Laptev, Amizadeh, and Flint]{laptev2015generic}
Nikolay Laptev, Saeed Amizadeh, and Ian Flint.
\newblock Generic and scalable framework for automated time-series anomaly detection.
\newblock In \emph{Proceedings of the 21th ACM SIGKDD international conference on knowledge discovery and data mining}, pages 1939--1947, 2015.

\bibitem[Lavin and Ahmad(2015)]{lavin2015evaluating}
Alexander Lavin and Subutai Ahmad.
\newblock Evaluating real-time anomaly detection algorithms--the numenta anomaly benchmark.
\newblock In \emph{2015 IEEE 14th international conference on machine learning and applications (ICMLA)}, pages 38--44. IEEE, 2015.

\end{thebibliography}

\newpage
\section{Appendix}
\subsection{Comparison with AttributionScanner}

In this section, we clarify the technical novelty of \systemname{}~in comparison to AttributionScanner~\cite{xuan2024attributionscanner}.  
\systemname{}~provides a problem-driven integration and adaptation designed for time series model validation—a domain fundamentally different from the image classification context of AttributionScanner. The novel aspects of \systemname{}~in comparison to AttributionScanner are summarized below.
\begin{itemize}[leftmargin=*]

    \item Domain differences: time series vs. image data.
    
    AttributionScanner is designed for image data, and its techniques are not directly applicable to solve the unique challenges of time series data. (1) AttributionScanner utilizes weighted data features to obtain data subgroups. However, unlike image data, which contains rich and spatially coherent semantics, time series data offers more implicit and less structured information, limiting the effectiveness of direct clustering based on data features. In \systemname{}, we design a novel similarity aggregation matrix to explicitly quantify the trivial differences between individual time sequences and their model attributes. (2) Time series are sequential and context-dependent, and similar patterns may occur at different time positions across instances. This means that similarity comparisons must account for temporal misalignment and local variations, which are not present in spatially static image inputs.
    Core components of our framework, such as DTW-based alignment, attribution over time steps, and behavior summarization of sequence clusters, were designed specifically to address the needs of sequential model interpretation. 
    
    \item Behavior summarization: temporal data and attribution similarities.
    
    \systemname{}~integrates temporal alignment (via DTW) with attribution similarity from CAM to support behavior summarization. This formulation is tailored to model behavior and data patterns for time series, and goes beyond feature-based clustering of AttributionScanner. Our design ensures that more nuanced differences between instances can be captured, and the resulting clusters are semantically and temporally aligned.
    
    \item Spuriousness propagation vs. label spreading. 
    
    While AttributionScanner and \systemname{}~operate at the cluster (or slice) level, their propagation mechanisms differ in intent and implementation.

   (1) \systemname{}~uses the label propagation algorithm, while AttributionScanner adopts a different algorithm, label spreading. We intentionally choose label propagation considering the unique characteristics of time series data. In contrast to image datasets, time series are often much smaller in scale and involve larger differences between different data groups. The label propagation algorithm is a closed-form solution and won’t change the human-labeled values throughout the process. While the label spreading used in AttributionScanner would vary the labeled data during interactions to tolerate potential noise, which may be suitable for large-scale image datasets, but introduces unwanted noise for our use scenario.
   
   (2) AttributionScanner uses label spreading over a slice similarity matrix, relying solely on feature-space similarity between slices to estimate spuriousness scores. In contrast, \systemname{}~performs semantic aggregation–based propagation: for each behavior cluster, we compare both its temporal pattern and model attribution profile against annotated clusters, and interpolate spuriousness scores based on these explicit behavior-level similarities. This approach avoids relying on latent feature similarity alone.
   
   % We have revised Section 3.4 to clarify this distinction and appropriately acknowledge AttributionScanner’s propagation strategy

   \item Problem-driven design and integration. 
   
   Each component of \systemname{}~was co-designed to support interpretability, rapid annotation, and model refinement in sequential data—from attribution-guided clustering, to cluster-based summaries, to propagation and retraining. This problem-specific integration forms an end-to-end pipeline for model debugging in time series.
   Our goal is not to present a new algorithmic primitive, but rather a system that extends human-in-the-loop validation to time series anomaly detection.
\end{itemize}

\begin{table}[b!]
\centering
\resizebox{\columnwidth}{!}{%
\begin{tabular}{p{3cm}<{\centering}p{3cm}<{\centering}p{3cm}<{\centering}p{3cm}<{\centering}}
\toprule[0.8pt]
KPI ID           & \#Points & Anomaly ratio & Time span (\#days) \\ \hline
02e99bd4f6cfb33f & 241189   & 4.37\%        & 183                \\
18fbb1d5a5dc099d & 240299   & 3.27\%        & 183                \\
1c35dbf57f55f5e4 & 240969   & 3.97\%        & 183                \\
da403e4e3f87c9e0 & 241148   & 3.17\%        & 183                \\ \bottomrule[0.8pt]
\end{tabular}%
}
\caption{Overview of KPI anomaly detection dataset for AIOps~\cite{li2022constructing}.}
\label{tab: AIOps}
\end{table}
\subsection{KPI anomaly detection dataset}

The KPI dataset~\cite{li2022constructing} is a real-world benchmark collected from multiple large-scale internet companies to support research on time series anomaly detection in AIOps (Artificial Intelligence for IT Operations). It includes 27 KPI categories with diverse patterns and anomaly types not typically captured by other public datasets~\cite{laptev2015generic,lavin2015evaluating}. Each timestamp is manually annotated with anomaly labels by experienced engineers.
For our evaluation, we select four categories with the highest anomaly ratios (see Table~\ref{tab: AIOps} for statistics). From each category, we extract equal-length sequences of 800 time points, resulting in a total of 1,200 time series instances.
%%%%%%%%%%%%%%%%%%%%%%%%%%%%% appendix end %%%%%%%%%%%%%%%%%%%%%%%%%%%%%%%

% %%%%%%%%%%%%%%%%%%%%%%%%%%%%% main paper start 2 %%%%%%%%%%%%%%%%%%%%%%%%%%%%%%%

% \newpage
% \section*{NeurIPS Paper Checklist}
% \input{Sections/06_checklist}

%%%%%%%%%%%%%%%%%%%%%%%%%%%%% main paper end %%%%%%%%%%%%%%%%%%%%%%%%%%%%%%%

\end{document}